\documentclass[12pt]{article}
\usepackage[nonatbib]{arxiv}
\usepackage[utf8]{inputenc}
\usepackage[T1]{fontenc}
\usepackage[numbers,comma,square,sort&compress]{natbib}
\usepackage{doi}
\usepackage{hyperref}
\hypersetup{pdfborder={0 0 0},
            colorlinks=true,
            urlcolor=blue,
            breaklinks=true,
            extension = }
\usepackage[document]{ragged2e}
\usepackage{graphicx}
\usepackage{enumerate}
\usepackage{amsmath}
\usepackage{amssymb}
\usepackage{xcolor}
\usepackage{textcomp}
\definecolor{comment}{RGB}{0,128,0} 
\definecolor{string}{RGB}{255,0,0}  
\definecolor{keyword}{RGB}{0,0,255} 
\usepackage{listings}
\lstdefinestyle{pseudo}{
    mathescape=true,
    numberstyle=\tiny,
    numbersep=5pt,
    frame=lines,
    captionpos=t,
    commentstyle=\color{comment},
    keywordstyle=\color{keyword},
    emph={volume_at,distance_to,physics_distance,get_geodetic},
    emphstyle=\color{red},
    emph={[2]volume,next,above,below},
    emphstyle={[2]\underbar},
    literate={:=}{{$\gets\ $}}1,
    morekeywords={Input,Function,begin,then},
    tabsize=4,
    breaklines=true,
    basicstyle=\footnotesize\ttfamily,
}
\newcounter{pseudocode}
\lstnewenvironment{pseudocode}[1][]{

    \setcounter{lstlisting}{\value{pseudocode}}
    \lstset{#1}
}{\addtocounter{pseudocode}{1}}
\lstdefinestyle{c}{
    commentstyle=\color{comment},
    stringstyle=\color{string},
    keywordstyle=\color{keyword},
    basicstyle=\footnotesize\ttfamily,
    numberstyle=\tiny,
    numbersep=5pt,
    frame=lines,
    breaklines=true,
    prebreak=\raisebox{0ex}[0ex][0ex]{\ensuremath{\hookleftarrow}},
    showstringspaces=false,
    upquote=true,
    tabsize=2,
}
\newcounter{program}
\lstnewenvironment{program}[1][]{

    \setcounter{lstlisting}{\value{program}}
    \lstset{#1}
}{\addtocounter{program}{1}}
\newcommand\ips[1]{\lstinline[language=Matlab, style=pseudo, columns=fixed,
    breaklines=false]{#1}}
\newcommand\ics[1]{\lstinline[language=c, style=C, columns=fixed,
    breaklines=false]{#1}}
\newcommand{\turtledef}[2]{%
        \href{https://niess.github.io/turtle-docs/\#v0.6/#1}{\ics{#2}}}
\newcommand{\turtlefun}[2]{%
        \href{https://niess.github.io/turtle-docs/\#v0.6/#1/#2}{\ics{#2}}}
\usepackage{float}
\usepackage[caption=false]{subfig}
\usepackage{caption}
\usepackage{multirow}

\usepackage[title]{appendix}
\appendixtitleon
\usepackage{authblk}

\usepackage{titlesec}
\titlelabel{\thetitle.\quad}
\usepackage{xwatermark}

\title{TURTLE: a C library for an optimistic stepping through a
topography}

\author[a\thanks{\tt{niess@in2p3.fr}}]{Valentin Niess}
\author[b]{Anne Barnoud}
\author[a]{Cristina Cârloganu}
\author[c]{Olivier Martineau-Huynh} 
\affil[a]{Université Clermont Auvergne, CNRS/IN2P3, LPC, F-63000
        Clermont-Ferrand, France}
\affil[b]{Université Clermont Auvergne, CNRS, IRD, OPGC, Laboratoire Magmas et
        Volcans, F-63000 Clermont-Ferrand, France}
\affil[c]{Sorbonne Université, Université Paris Diderot, Sorbonne Paris Cité,
        CNRS, Laboratoire de Physique Nucléaire et de Hautes Energies (LPNHE),
        4 place Jussieu, F-75252, Paris Cedex 5, France}

\begin{document}
\maketitle

\noindent\rule{\textwidth}{0.4pt}
\vskip 0.05in
\begin{abstract}
\justify
TURTLE is a C library providing utilities allowing to navigate through a
topography described by a Digital Elevation Model (DEM). The library has been
primarily designed for the Monte~Carlo transport of particles scattering over
medium to long ranges, e.g.~atmospheric muons. But, it can also efficiently
handle ray tracing problems with very large DEMs ($10^9$ nodes or more),
e.g.~for neutrino simulations. The TURTLE library was built on an
\emph{optimistic} ray tracing algorithm, detailed in the present paper. This
algorithm proceeds by trials and errors, approximating the topography within the
modelling uncertainties of the DEM data. This allows to traverse a topography in
constant time, i.e.~independently of the number of grid nodes, and with no added
memory.  Detailed performance studies are provided by comparison to other ray
tracing algorithms and as an application to muon transport in a Monte~Carlo
simulation.

\end{abstract}

\noindent\rule{\textwidth}{0.4pt}

\justify
\section{Introduction}

Monte~Carlo (MC) simulations are a key element of many physics experiments. Some
transport problems require to propagate particles over tens to hundreds of km,
in a complex environment.  This is for example the case for muography
measurements or for experiments searching extraterrestrial neutrinos.  Muography
is a particularly complex case since particles might scatter, e.g.~by grazing
the ground and interacting with surface structures. Accurate Monte~Carlo
computations require a detailed description of the topography over large
distances.

Topography data are provided by a Digital Elevation Model (DEM), which can also
be a combination of several DEMs, e.g.~with different resolutions. A DEM can
represent various information, e.g.~the surface above ground structures and the
canopy, the ground level, the sea level (geoid undulations) w.r.t.~a reference
ellipsoid. The elevation data, $z$, are commonly stored over a regular grid in
geodetic coordinates (latitude, longitude) or projected coordinates, $(x, y)$,
e.g.~Universal Transverse Mercator (UTM).  Such DEMs generate simple surfaces:
$z=f(x, y)$.  Note that complex features, e.g.~cavities or overhanging cliffs,
would map several elevation values to a same projected coordinate, $(x, y)$. By
definition, the simple DEMs considered here cannot contain such topologies.
Unstructured meshes are neither considered in the following. Elevation values
are typically stored using 2 bytes (\ics{int16_t}) per grid node.  Large scale
data sets comprise $10^8$ to $10^9$ nodes resulting in a hundreds of MB to a few
GB memory footprint. Due to the Earth curvature, when transformed to the
Cartesian laboratory frame of the simulation, the grid nodes are no longer
equally spaced.  Planarity is lost as well over large distances. For example two
nodes at the same elevation above sea level, but distant by 10\,km, are actually
vertically separated by about 8\,m in the laboratory frame.

In MC simulations, a geometry is defined as a collection of closed volumes,
bounded by surfaces, and filled with various materials. The transport through
the geometry is simulated by discrete steps joining interaction vertices along
straight line segments.  Each Monte~Carlo step involves a ray tracing problem.
Ray tracing is a generic geometry problem. For example, it is also encountered
when rendering high quality 3D scenes in computer graphics, since it allows to
account for physical effects like light reflection and refraction.  However, in
a MC the ray length can be drastically limited by the physics, as interactions
occur with the matter of volumes or with external electromagnetic fields.  This
leads to changes in the particle direction of propagation, at each interaction
vertex, requiring specific ray tracing optimisations.  The
pseudocode~\ref{alg:transport} describes a simple transport algorithm in order
to guide the following discussion.

\begin{pseudocode}[language=Matlab,style=pseudo,caption=Simple Monte~Carlo
        transport algorithm, label=alg:transport,float=ht!]
Input: $\vec{r}, \vec{u}$ % The particle position and direction

begin
    % Get the initial volume
    volume := volume_at($\vec{r}$)

    % Iterate until something happens
    while $\ldots$
        % Get the step limit from the physics
        $\lambda$ := physics_distance(volume, $\vec{r}$, $\ldots$)

        % Get the distance to the next volume, or an underestimate
        $s$, next := distance_to(volume, $\vec{r}$, $\vec{u}$)

        if $s \geq \lambda$ then
            % Limit the step according to the physics
            $s$ := $\lambda$
        else
            % Update the volume
            if next then
                volume := next
            else
                volume := volume_at($\vec{r}$)
            end
        end

        % Update the particle properties
        $\vec{r}$ := $\vec{r} + s \vec{u}$
        $\ldots$
    end
end
\end{pseudocode}

A MC transport engine navigates through the geometry using the pseudocode
functions: \ips{volume_at} and \ips{distance_to}.  The former function returns a
reference to the volume at a given position, $\vec{r}$. The latter one provides
the distance to the closest boundary surface, along a direction, $\vec{u}$. When
volumes are explicitly connected, the next volume can also be returned. Some
Monte~Carlo engines, like MCNP\,\cite{MCNP:5,MCNPX,MCNP:6.2}, enforce this when
defining the geometry.  Otherwise the next volume must be found using the
\ips{volume_at} function. Note that an underestimate for the distance to the
next volume can be returned instead of the exact one. The pseudocode expects a
reference to the current volume to be returned in this case.  Underestimates,
independent of the particle direction, can be significantly faster to compute
even for usual geometric shapes: cube, sphere, $\ldots$ In the case where the
Monte~Carlo navigation is limited by the physics, $\lambda < s$ for most steps,
then using underestimates may speed up the navigation. This strategy is used by
the Geant4\,\cite{Geant4:2003,Geant4:2006, Geant4:2016} multi-purpose MC
package.  Geant4 is a popular Monte~Carlo engine for particle physics. It was
originally designed for modelling the interactions of high energy particles
within particle physics detectors. The Geant4 engine is flexible though, and
there have been many other applications, e.g.~in medical physics or for space
radiation studies.

In the following, for the sake of clarity, we consider a simple geometry with
only two volumes: the atmosphere and the Earth, separated by a ground surface.
The ground surface is usually approximated as a collection of plane facets
joining the DEM nodes.  Tessellating a regular grid of $n$ nodes with triangles
requires approximately two triangles per node in the large $n$ limit.
Efficiently building a 3D model for a non parametric volume containing billions
of facets is rather involved.  One should carefully control the memory overhead,
i.e.~the extra memory used by the 3D model in addition to the original DEM data.
For example, adding a single precision float (32~bits) per node already triples
the memory usage. At the same time, the geometry must be smartly organised in
order to speed up intersection searches.

A generic optimisation method is to sort geometric shapes (volumes, surfaces,
$\ldots$) using a Bounding Volume Hierarchy (BVH). The shapes  are organised in
a tree structure according to their bounding boxes. The tree is sorted in order
to speed up subsequent geometry operations, e.g.~computing the intersection of
the shapes with a line or with a plane. BVHs are used in 3D modelling for ray
tracing or collision detection.

Optimising the MC navigation for volumes bounded by a tessellated surface, or by
3D meshes, is a problem that has been previously investigated, e.g.~in Geant4
for medical physics applications. In particular, it was pointed out by
\citet{Poole:2012} that for physics limited navigation, a polyhedral meshing of
a volume interior can be more efficient than tessellating its bounding surface.
Dividing the interior of a volume in virtual sub-volumes naturally provides fast
underestimates of the distance to the next boundary.  Unfortunately, this study
was done at a time where BVH optimisation was not implemented for tessellated
surfaces (\ics{G4TessellatedSolid}). BVH acceleration (\ics{SmartVoxels}) was
however used for the polyhedral mesh. In recent versions of Geant4 (e.g.~10.5)
the VecGeom library~\cite{VecGeom:site} can be used to speed up tessellated
geometries by using a SIMD optimized BVH, see e.g.~\citet{VecGeom:proceeding}).
Therefore, the conclusion of \citet{Poole:2012} could be different today.

For regular DEM grids, the memory overhead due to the geometry modelling can be
kept down to zero. The geometry can be fully computed on the fly from the
initial elevation data. In the case of a regular polyhedral mesh, there exists a
natural alternative to BVH trees with zero memory cost: the nodes connectivity.
In this case, the exit facet from a polyhedron determines the next volume on the
grid.  This allows to traverse the full grid in $\mathcal{O}(\sqrt{n})$
operations, $n$ being the number of nodes. In comparison, a BVH tree could in
principle perform the same in $\mathcal{O}(\log(n))$ operations. Nevertheless,
the former algorithm costs no extra memory when the grid is regular. It could be
more efficient when the navigation is limited by the physics.

In addition to the previous issues, let us point out that approximating a
topographic surface with flat triangular facets between the nodes is a poor man
solution.  Instead, higher degree interpolations could be used. With additional
developments one might replace the flat facets by smooth surfaces between the
nodes, e.g.~splines or NURBS.  However, we propose a much simpler and
straightforward solution to overcome the discussed issues of memory usage,
stepping efficiency and accuracy. This solution is called \emph{optimistic}
algorithm in the following.  It implies computing an approximate distance to the
next volume, rather than an exact one. The compromise of using an approximate
stepping can be motivated by the fact that elevation values provided by a DEM
are actually space averages around the node coordinates. The typical accuracies
of such models is of the order of 10\,cm (see e.g.~\citet{MUKHERJEE2013205}).
The \emph{optimistic} algorithm is presented in next
section~\ref{sec:algorithm}. It is the baseline of the TURTLE library described
in section~\ref{sec:library}.  Test cases are presented in
section~\ref{sec:benches}. Those are used in order to tune the \emph{optimistic}
algorithm in section~\ref{sec:balancing}. Then, we provide detailed performance
studies by comparison to other ray tracing algorithms
(section~\ref{sec:comparison}) and as an application to muon transport in a
Monte~Carlo simulation (section~\ref{sec:hepmc}).

\section{The \emph{optimistic} stepping algorithm \label{sec:algorithm}}

An approximate solution to the ray tracing problem is to proceed by trials and
errors. When navigating through the DEM data, a tentative (\emph{optimistic})
stepping distance is proposed for the topography, at each Monte~Carlo step.  If
the step leads to a ground crossing, the boundary is located with a binary
search.  The stepping distance is reduced iteratively in order to converge to
the boundary within an accuracy of $\epsilon$.  If there is no ground crossing
the step is accepted.  Note that if the initial step length, $s_0$, is too long
one might overrun some details of the topography of size smaller than $s_0$.
Consequently, this method has a non-null failure rate.  Setting a small enough
constant value, e.g.~$s_0 = 1\,\mathrm{cm}$, mitigates the risk but is
very inefficient.

Yet, in the case of a ground surface described by a DEM, a simple yet efficient
guess for the initial stepping distance, $s_0$, is:
\begin{equation} \label{eq:step_size}
s_0 = \max\left(\alpha | h_0 - g_0 |,\,s_\mathrm{min}\right)
\end{equation}
where $h_0$ is the current altitude of the particle and $g_0$ the corresponding
ground level, i.e.~$| h_0 - g_0 |$ is the difference in height w.r.t.~the
ground. The parameters $\alpha \leq 1$ and $s_\mathrm{min}$, in
eq.~(\ref{eq:step_size}), are two tuning factors. Let us call them slope factor
and resolution factor in the following.

Note that the altitudes $h_0$ and $g_0$ in equation~\ref{eq:step_size} are
considered in the same local referential, i.e.~along the local vertical.  This
is important for particles propagating over long ranges. It ensures that the
algorithm properties are not affected by the Earth curvature. Usually the ground
altitude w.r.t.~sea level, $g_0$, can be directly read from DEMs, given the
particle geodetic coordinates, i.e.~latitude and longitude. A Monte~Carlo would
however operate in a fixed Cartesian frame instead, e.g.~a geocentric frame.
This implies extra computations for transforming the particle coordinates at
each step.

Using the guess provided by equation\ \ref{eq:step_size} instead of a constant
step length can provide impressive speed-up, by several orders of magnitude as
shown in section~\ref{sec:balancing}, without significant loss of accuracy. The
driving idea is that as long as slopes are not too steep, the height w.r.t~the
ground is a safe estimate of the closest distance to the ground.  Reducing the
slope factor allows to handle steeper and stepper slopes. Note that setting
$\alpha$ below $1$ is not efficient for vertical trajectories.  In this case, an
improvement to equation~\ref{eq:step_size} could be to vary $\alpha$ depending
on the particle direction w.r.t.~the local vertical. Yet, in this paper we
consider only close to horizontal trajectories. The resolution factor plays the
role of a safeguard against numerical errors, e.g.~close to a boundary. While
the boundary is approached with an accurate binary search, it is departed from
using a constant step size, given by the resolution factor.

Below are pseudocode implementations of the \ips{volume_at} and
\ips{distance_to} functions, using the \emph{optimistic} algorithm and
eq.~(\ref{eq:step_size}) as initial guess for the step length. A key component
of the method is the \ips{get_geodetic} function. Given a particle position in
the laboratory frame, this function computes the altitude and the corresponding
ground level with respect to a common reference.  This latter data is
efficiently extracted from the DEMs, on the fly, i.e.~without building a
complete model of the ground surface. This is the main purpose of the TURTLE
library. As a consequence, a Monte~Carlo transport engine may navigate quickly
and simply through the topography using the \emph{optimistic} approach.
\\
\begin{pseudocode}[language=Matlab,style=pseudo,caption=Example of
    \ips{volume\_at} function for an \emph{optimistic} stepping through
    topography data, label=alg:optimistic:1]
Input: $\vec{r}$ % The particle position

Function volume_at($\vec{r}$)
    $h, g$ := get_geodetic($\vec{r}$)
    if $h > g$ then
        return above
    else
        return below
    end
end
\end{pseudocode}

\newpage

\begin{pseudocode}[language=Matlab,style=pseudo,caption=Example of
    \ips{distance\_to} function for an \emph{optimistic} stepping through
    topography data, label=alg:optimistic:2]
Input: $\vec{r}, \vec{u}$       % The particle position and direction
       volume % A reference to the current volume

Function distance_to(volume, $\vec{r}$, $\vec{u}$)
    % Compute the tentative step length
    $h_0, g_0$ := $\ $get_geodetic($\vec{r}$)
    $s_0$ := $\ \max\left(\alpha | h_0 - g_0 |,\,s_\mathrm{min}\right)$

    % Check the tentative position
    next := volume_at($\vec{r} + s_0 \vec{u}$)

    if next $\neq$ volume then
        % Locate the topography crossing with a binary search
        $s_1$ := $\ 0$
        while $s_0 - s_1 > \epsilon$
            $s_2$ := $\ (s_0 + s_1) / 2$
            volume := volume_at($\vec{r} + s_2 \vec{u}$)
            if volume = next then
                $s_0$ := $\ s_2$
            else
                $s_1$ := $\ s_2$
            end
        end
    end

    return $s_0$, next
end
\end{pseudocode}

Note that it would be enough to return the signed distance to the ground, $h -
g$, from the \ips{get_geodetic} function.  However, providing the particle and
ground altitude allows to extend the algorithm by adding volumes bounded by
additional topography surfaces. For example, one can define a seawater volume
bounded by $g < h \leq 0$, with altitudes measured w.r.t.~the sea level. An
atmosphere can be defined on top of that as $\max(g, 0) < h < h_\mathrm{max}$,
with $h_\mathrm{max}$ the top of the atmosphere. Similarly one could specify an
inner structure of the Earth, e.g.~following the Preliminary Reference Earth
Model (PREM) of~\citet{PREM}. When multiple topography surfaces are used, the
algorithm discussed in this paper must be slightly modified. An initial guess,
$s_0$, must be computed for both the surface above and below the current volume,
using eq.~(\ref{eq:step_size}). Then the smallest value is used for the
tentative step.

\section{The TURTLE library \label{sec:library}}

\textbf{T}opographic \textbf{U}tilities for t\textbf{R}ansporting
par\textbf{T}icules over \textbf{L}ong rang\textbf{E}s (TURTLE) is a C library
providing utilities for stepping through a topography described by a DEM, using
the \emph{optimistic} algorithm, described in section~\ref{sec:algorithm}. The
source code of the library is available from GitHub\,\cite{TURTLE:GitHub} under
the GNU LGPL-3.0 license. Version 0.7 is used in this paper. The library was
written in C99 and unit-tested on Linux and OSX with a coverage of $90\,\%$. A
\texttt{Makefile} and a \texttt{CMakeLists.txt} are shipped with the source code
allowing to build TURTLE as a shared library, as well as some examples. Note
that in addition to the standard C library, TURTLE also requires
\texttt{LibTIFF} and \texttt{libpng} in order to read GeoTIFF and PNG files
respectively. Though, these functionalities can be disabled with macro
definitions when compiling the library, removing the corresponding dependencies.

TURTLE is neither an image processing library nor a Monte~Carlo transport
engine.  It focuses on few functionalities in order to efficiently track
particles over large ranges. These functionalities are exposed to the end user
by following an Object Oriented (OO) design, as can be seen by browsing the
documentation of the Application Programming Interface
(API)\,\cite{TURTLE:docs}.

\subsection{Maps and projections}

The base object of the TURTLE library is an opaque \turtledef{group/map}{struct
turtle_map} object. It encapsulates a DEM.  Those can be loaded from data
files with the \turtlefun{group/map}{turtle_map_load} function. Note that TURTLE
can only load a few commonly used data formats for geographic maps: GEOTIFF
(e.g.~used by ASTER\,\cite{ASTER}), \texttt{*.hgt}~(e.g.~used by
SRTMGL1\,\cite{SRTMGL1}) or \texttt{*.grd}~(e.g.~used by EGM96\,\cite{EGM96}).
Image formats must be 16\,bits and grey-scale.  New maps can also be created
empty and filled using the \turtlefun{group/map}{turtle_map_create} and
\turtlefun{group/map}{turtle_map_fill} functions. This allows to create custom
readers for example. In addition, TURTLE supports dumping and loading maps in
PNG, enriched with a dedicated header as a \texttt{tEXt} chunk.

The elevation at any position is computed from the DEM with the
\turtlefun{group/map}{turtle_map_elevation} function. A bilinear interpolation
is used in order to estimate the elevation between DEM nodes. The bilinear
interpolant $\hat{f}(x, y)$ of $f$ over $[x_i, x_{i+1}] \times [y_j, y_{j+1}]$
can be written as:
\begin{align} \label{eq:bilinear}
        \hat{f}(x, y) & = \overline{h}_x \overline{h}_y f_{i,j} +
                    h_x \overline{h}_y f_{i + 1,j} +
                    \overline{h}_x h_y f_{i, j + 1} +
                    h_x h_y f_{i +1, j + 1} \\
        h_x & = \frac{x - x_i}{x_{i+1} - x_i}, \overline{h}_x = 1 - h_x \\
        h_y & = \frac{y - y_j}{y_{j+1} - y_j}, \overline{h}_y = 1 - h_y \\
        f_{i,j} & = f(x_i, y_j)
\end{align}
where $x_i$, $x_{i+1}$, $y_j$ and $y_{j+1}$ are the nodes surrounding $(x, y)$.
Note that contrary to a triangular tessellation, this interpolant is quadratic
w.r.t.~the position $(x, y)$, producing some smoothing. The interpolated values
are continuous at the node boundaries. However, the first derivative is not
resulting in ridges at the borders linking two nodes. Bilinear interpolation was
chosen over triangular tessellation or bi-cubic interpolation as a compromise
between speed and accuracy.  With this approach a map keeps in memory no more
than the initial elevation data, stored using 2 bytes per node.  The terrain is
modelled on the fly when an elevation value is requested. The elevation values
at nodes can also be directly inspected using the
\turtlefun{group/map}{turtle_map_node} function.  The map meta data can be
retrieved with the \turtlefun{group/map}{turtle_map_meta} function, e.g.~the
size of the grid.

Local maps, e.g.~UTM projections, have an associated opaque
\turtledef{group/projection}{struct turtle_projection} object. This object
allows to convert between the local map coordinates and the geodetic ones, i.e.
latitude and longitude. This is done with the
\turtlefun{group/projection}{turtle_projection_project} and
\turtlefun{group/projection}{turtle_projection_unproject} functions. The
\turtlefun{group/map}{turtle_map_projection} function allows to borrow a pointer
to the map projection. \ics{NULL} is returned if the map coordinates are
geodetic ones. A projection object can also be directly created (destroyed)
using the \turtlefun{group/projection}{turtle_projection_create}
(\turtlefun{group/projection}{turtle_projection_destroy}). Note that the
projection borrowed from a map should not be explicitly destroyed. It is
automatically destroyed when calling \turtlefun{group/map}{turtle_map_destroy}.
A projection can be modified with the
\turtlefun{group/projection}{turtle_projection_configure} function. Its name
is provided by the \turtlefun{group/projection}{turtle_projection_name}
function. Note that when a map is dumped in PNG format, the projection name
is also written to the file.

Two types of projections are available: Lambert Conformal Conic (LCC) and
Universal Transverse Mercator (UTM). See e.g.~\citet{WILLIAMS1995} for a
description of these projections. The type and parameters of a projection are
encoded as a name string at its creation, e.g.~\ics{"UTM 31N"} for a UTM
projection using zone 31 of the northern hemisphere or \ics{"UTM 31S"} for the
southern hemisphere. Note that we use the simplified notation for UTM zones,
specifying the hemisphere as \ics{'N'} or \ics{'S'} instead of the latitude
band. Alternatively, one can also directly specify the central meridian
($\lambda_0$) instead of the zone followed by the hemisphere, e.g.~\ics{"UTM
3.0N"} for $\lambda_0 = 3\,^\mathrm{o}$ East. For LCC projections only a few
zones are currently available, corresponding to some sets of parameters commonly
used in France, see e.g.~IGN documents\ \cite{IGN:projections}.
Table~\ref{tab:projections} provides a summary of the supported projections and
of their corresponding encoding.

\begin{table}[ht]
\begin{center}
\begin{tabular}{|ll|} \hline
    Projection           & Format                \\
\hline \hline
    Lambert I            & {\ics{"Lambert I"}}   \\ 
    Lambert II           & {\ics{"Lambert II"}}  \\
    Lambert II extended  & {\ics{"Lambert IIe"}} \\
    Lambert III          & {\ics{"Lambert III"}} \\
    Lambert IV           & {\ics{"Lambert IV"}}  \\
    Lambert 93           & {\ics{"Lambert 93"}}  \\
    \multirow{2}{*}{UTM} & {\ics{"UTM \%d\%c"}}  \\
                         & {\ics{"UTM \%lf\%c"}} \\ \hline
\end{tabular}
\caption{List of supported projections in TURTLE \texttt{v0.7}. The format of
        the name string is indicated following the \ics{printf} semantic.
        \label{tab:projections}}
\end{center}
\end{table}

\subsection{Stacks and clients}

Worldwide models, e.g.~ASTER\,\cite{ASTER} or SRTMGL1\,\cite{SRTMGL1} are
usually divided in tiles, e.g.~of $1\,^\mathrm{o} \times 1\,^\mathrm{o}$. These
models are encapsulated  in TURTLE as a \turtledef{group/stack}{struct
turtle_stack} of maps, all maps using geodetic coordinates. When a ground
elevation is requested, using \turtlefun{group/stack}{turtle_stack_elevation},
the stack automatically manages loading the right tile into memory. Tiles are
kept in memory until the maximum stack size is reached. In this case the oldest
accessed map is removed from memory in order to make room for a newly loaded
one. The stack can also be manually cleared with the
\turtlefun{group/stack}{turtle_stack_clear} functions.  The maximum stack size
is specified at the stack creation, using the
\turtlefun{group/stack}{turtle_stack_create} functions. Providing a negative or
null value disables the automatic deletion of old maps, resulting in all maps
being kept in memory, as they are loaded. In some cases it might be more
efficient to load all maps in memory right from the start. This is achieved with
the \turtlefun{group/stack}{turtle_stack_load} function. At the stack creation
one can also provide a couple of
\turtlefun{group/callback}{turtle_stack_locker_t} callbacks. These callbacks
must provide a lock/unlock mechanism for exclusive data accesses in
multithreaded usage, e.g.~by managing a semaphore.

Accessing the elevation data directly from a stack object is not thread safe.
Instead, for multithreaded usage one must instantiate a
\turtledef{group/client}{struct turtle_client} of the stack, one per thread.
Note that the targeted stack must have lock and unlock callbacks, otherwise an
error is generated. The maps usage is monitored by the stack by reference
counting. Each client holds a reference to one map at most. When an elevation
value is requested, with the \turtlefun{group/client}{turtle_client_elevation}
function, the client first checks if the request belongs to its current map. If
not, the client interacts with the stack in order to get access to the right
map. The reference to any previous map is dropped as well. Maps with at least
one reference cannot be dropped by the stack, even though the stack size is
exceeded. Note that switching the client's map requires to lock the stack.
However, since we are tracking particles, most of successive elevation accesses
are expected to fall within the same map. Therefore, these locks generate no
significant retention.

\subsection{ECEF coordinates transforms}

Elevation maps are provided in geodetic coordinates or using a local projection.
Projected coordinates cannot be used directly in a large scale Monte~Carlo where
the Earth curvature is no more negligible. A convenient and correct reference
frame to use in such Monte~Carlo simulations is the Earth-Centered Earth-Fixed
(ECEF) frame. The \turtledef{group/ecef}{turtle_ecef} functions provide
coordinate transforms from and to ECEF.

The geodetic to ECEF transform
(\turtlefun{group/ecef}{turtle_ecef_from_geodetic}) is straightforward, similar
to spherical to Cartesian coordinates. The only difference is an ellipticity
factor due to the fact that geodetic coordinates are given w.r.t.~a reference
ellipsoid (WGS84 typically), not a sphere. The reverse transform
(\turtlefun{group/ecef}{turtle_ecef_to_geodetic}) is less trivial. We
implemented the closed form which is computed efficiently according
to~\citet{Olson:1996}.  Converting between projected map coordinates and ECEF
is performed via intermediate geodetic coordinates. It requires to project or
un-project the local coordinates to or from geodetic ones, such that the
\ics{turtle_ecef} functions can be used.

In order to define a direction it is often convenient to use local angular
coordinates as for example horizontal ones, i.e.~azimuth and elevation. The
\turtlefun{group/ecef}{turtle_ecef_to_horizontal} and
\turtlefun{group/ecef}{turtle_ecef_from_horizontal} provide such transforms.
Note that the geodetic longitude and latitude must be provided in both cases, in
order to define the local East, North, Up (ENU) frame used for the horizontal
coordinates.

\subsection{The stepper}

We now come to the higher level functionalities of TURTLE. The main purpose of
the library is to provide tools allowing a Monte~Carlo simulation to step
efficiently through a topography using the \emph{optimistic} approach and ECEF
coordinates.  The \turtledef{group/stepper}{struct turtle_stepper} object
provides an encapsulation of the topography data focused on Monte~Carlo
stepping. A \ics{turtle_stepper} is instantiated with the
\turtlefun{group/stepper}{turtle_stepper_create} function. It starts empty, i.e.
without any elevation data. DEMs can be added with the
\turtlefun{group/stepper}{turtle_stepper_add_map} and
\turtlefun{group/stepper}{turtle_stepper_add_stack} functions. These functions
allow to specify an offset to the native DEM elevation values. A flat topography
is specified with the \turtlefun{group/stepper}{turtle_stepper_add_flat}
function. The last added entry is on the top of the data stack. When an
elevation value is requested, the stepper scans its stack down, starting from
the top, and it returns the first matching data. This allows to implement
different levels of detail, depending on the area of interest. For example, a
typical usage would be to have an accurate local map as the top layer, followed
by a stack of geodetic maps from a world wide DEM, with a coarser resolution but
a larger coverage, and finally a constant altitude as a fallback for very large
distances, when data are missing.

Though not used in this paper, complex geometries can be defined as well by
using multiple topography layers. A topography layer can for example specify the
depth of the soil, or the level of (sea) water, or the height of the vegetation,
etc.  Adding a topography layer is done with the
\turtlefun{group/stepper}{turtle_stepper_add_layer} function. Then, multiple
data can be added to the new layer, as described previously, i.e.~with the
\turtlefun{group/stepper}{turtle_stepper_add_flat},
\turtlefun{group/stepper}{turtle_stepper_add_map} and
\turtlefun{group/stepper}{turtle_stepper_add_stack} functions. A single DEM can
be used in several layers, e.g.~with different offset values.

An additional difficulty might arise from the fact that elevation data are
usually provided w.r.t.~the geoid (sea level), not the ellipsoid. On very large
scales this might have an impact since the difference between the two can reach
hundred meters. In this case, when computing ECEF coordinates, one must
convert the elevations above the geoid to heights above the ellipsoid. This is
done automatically by the stepper by initially providing a map of geoid
undulations, e.g.~EGM96\,\cite{EGM96}, with the
\turtlefun{group/stepper}{turtle_stepper_geoid_add} function.

Once the \ics{turtle_stepper} data is configured, the
\turtlefun{group/stepper}{turtle_stepper_step} function allows to perform
elementary steps through the topography. It has two modes of operation. In both
modes, it takes as input an initial position in ECEF.
\begin{enumerate}[(i)]
        \item When no stepping direction is provided, the function returns the
                geodetic coordinates at the given ECEF position, the ground
                level and a tentative stepping distance based on
                eq.~(\ref{eq:step_size}).

        \item When a stepping direction is provided, the function performs a
                single step following function \ips{distance_to} described
                in pseudocode~\ref{alg:optimistic:2}. In this case, the
                function returns the geodetic coordinates at the final position,
                the corresponding ground level and the actual step length.
\end{enumerate}
Both modes rely on eq.~(\ref{eq:step_size}). The parameters $\alpha$ and
$s_\mathrm{min}$ can be modified with the
\turtlefun{group/stepper}{turtle_stepper_slope_set} and
\turtlefun{group/stepper}{turtle_stepper_resolution_set} functions. Mode (i)
is meant to be integrated in a MC. It provides only a stepping distance, but
it actually does not perform any stepping. On the other hand, mode (ii) can be
directly iterated in order to step through the topography.  However, it is
likely to be sub-optimal when integrated within a transport engine, because it
does not take into account physical processes nor other volumes, both of which
might limit the step length as well.

Computing the ground level at a given position in the laboratory frame requires
to convert the ECEF position to the geodetic coordinates used by maps, or to
projected ones, e.g.~UTM. Doing so can be costly CPU wise, especially when
performing a binary search. Therefore the \ics{turtle_stepper} object implements
several optimisations. The first one consists in recording any transformed
coordinates the first time that a conversion occurs. Similarly, the elevation
value of DEMs are recorded after their first access. Then, when the transform or
DEM is requested again during the same step, the result is read back from the
cache instead of being re-computed. In addition, the stepper also stores a
copy of the last outcome of a step, such that it does not need to be computed
again if the same position is requested successively. This allows to efficiently
chain calls to \turtlefun{group/stepper}{turtle_stepper_step}.

Another type of optimisation consists in computing a local linear approximation
(LLA) of the transform from the ECEF coordinates to the maps ones, centered on
the last computed result. Then, when new positions are requested close enough
from the LLA center, the LLA is used instead of the exact computation. The range
over which the LLA is used, $r_\mathrm{LLA}$, can be configured with the
\turtlefun{group/stepper}{turtle_stepper_range_set} function. Computing the LLA
has a non-negligible CPU cost. It is worth only if one can ensure that the
following queries actually use it. Therefore LLAs are computed only if the
current step length is smaller than $\frac{r_\mathrm{LLA}}{3}$.

\subsection{Error handling}

Last, but not least, errors in TURTLE are meant to be managed with a
\turtlefun{group/callback}{turtle_error_handler_t} callback. Whenever a
library function encounters an error this callback is called. It gets as input
an \turtledef{type/turtle\_return}{enum turtle_return} code and a
\turtlefun{group/callback}{turtle_function_t}, indicating the type of error
that occurred and the faulty function. In addition a brief description of the
error is also provided as a string.

The default behaviour is to print the brief error description to \ics{stderr}
and to \ics{exit} to the OS. This behaviour can be overridden be providing a
custom error handler with the
\turtlefun{group/error}{turtle_error_handler_set} function. Setting the
handler to \ics{NULL} disables error handling. Note that in this case, most
library functions return a \ics{turtle_return} code in order to indicate
their exit status, i.e.~\ics{TURTLE_RETURN_SUCCESS} on success, or an error
code otherwise.

\section{Test cases \label{sec:benches}}

In the following sections we present the results of various tests and
comparisons that have been carried out with the TURTLE library. Most of the
source code used for these tests can be found on
GitHub~\cite{TURTLE_PERFS:GitHub}.  The performance critical sections have been
implemented in C or C++. Higher level functionalities have been scripted using
LuaJIT~\cite{LuaJIT} and its \emph{foreign function interface} (\ics{ffi})
package. The test suite was build with CMake (\ics{"Release"} build) and
gcc~4.8.5, the system native compiler. It was run on a dedicated CentOS7 server
hosting 64 cores at 2.2\,GHz (Intel$^{\tiny{\textregistered}}$
Xeon$^{\tiny{\textregistered}}$ E5-4620) with 128\,GB of DDR3 memory. Two
applications were considered. The first one consists in computing the rock depth
along straight lines. Knowing the rock depth along a Line Of Sight (LOS) is
important for muography applications. The muon ($\mu$) flux transmitted through
the target indeed depends primarily on the integrated density along the line of
sight.  Combining the flux measurement with the rock depth thus provides an
estimate of the average density of the target. On larger scales, the rock depth
close to the horizon is also meaningful in order to estimate the target mass for
high energy neutrinos.  The first case allows to test the performances of the
\emph{optimistic} algorithm as a pure geometry solver. The second application
tests the performances of the TURTLE library when integrated in a Monte~Carlo.
It consists in computing the spectrum of transmitted atmospheric $\mu$ using the
PUMAS\,\cite{PUMAS:GitHub} MC engine.

\begin{figure}[!t]
  \centering
  \subfloat{\includegraphics[width=0.5\textwidth]{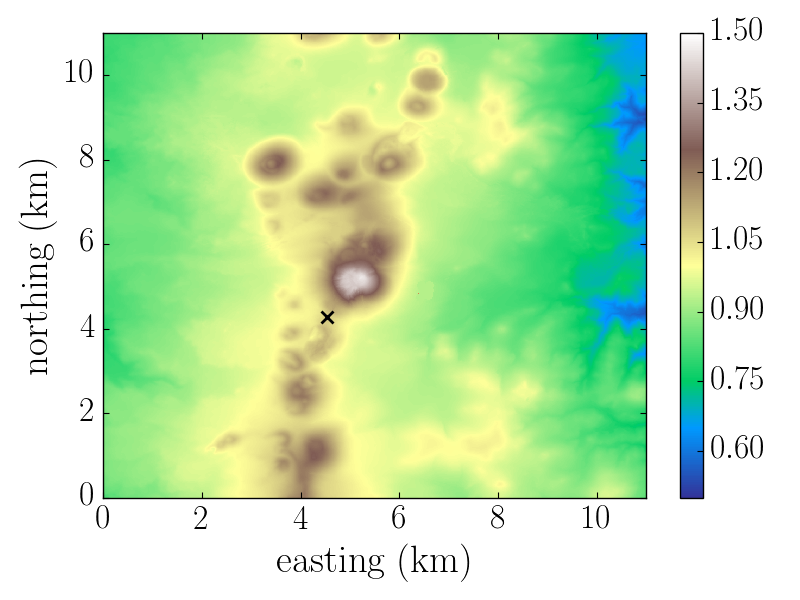}}
  \subfloat{\includegraphics[width=0.5\textwidth]{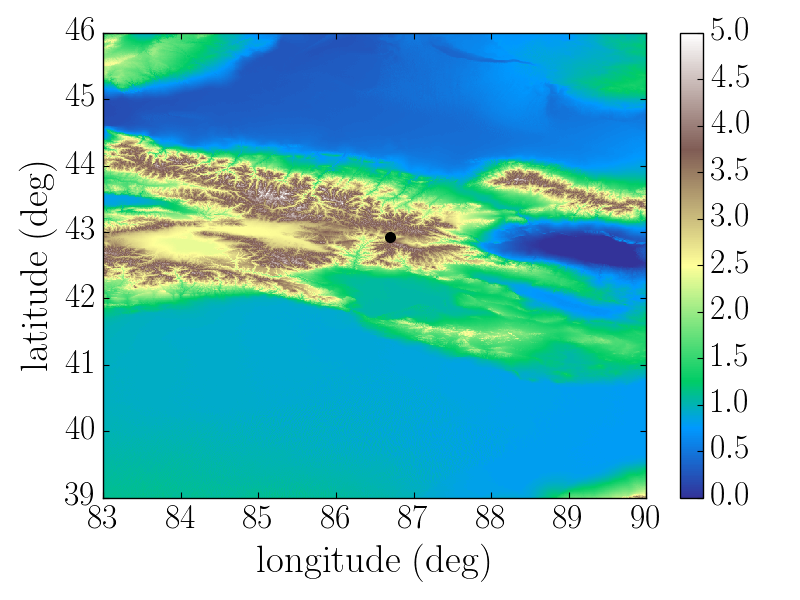}}
  \caption{Elevation maps (km) of the case studies. Left: local map of the
        Chaîne des Puys (Lambert 93 projection). The black cross indicates
        the CDC location. Right: Tian Shan mountains. The black dot in the
        middle of the map indicates the location of the Ulastai valley.
        \label{fig:map}}
\end{figure}

For both applications we consider two view points located in two mountainous
areas: Chaîne des Puys in France (fig.\,\ref{fig:map}, left) and Tian Shan in
China (fig.\,\ref{fig:map}, right).  These areas correspond to real experiments
with different topographies and data sets. The first area is taken from
muography data acquisition (\citet{ambrosino2015joint}) targeting the Puy de
Dôme volcano, highest summit (1\,465\,m) of the Chaîne des Puys. The Chaîne des
Puys is a north-south oriented chain of $\sim$80 volcanoes in the Massif Central
in France. We consider a view point where muography detectors have been
operated: Col de Ceyssat (CDC) located south-west of Puy de Dôme at
45.764160\,$^\mathrm{o}$\,N, 2.955385\,$^\mathrm{o}$\,E, 1\,080\,m. It is
indicated with a black cross on the left plot of fig.~\ref{fig:map}. For the
second area we select a view point in the Ulastai valley. It is a high altitude
valley (2\,650\,m high) in the Tian Shan mountain range. Compared to the Chaîne
des Puys this is a rocky area with high and steep summits. The Ulastai valley is
surrounded by 5\,000\,m high peaks. Ulastai is hosting the 21\,CMA\,\cite{21CMA}
and TREND\,\cite{Ardouin:2011} experiments.  The TREND experiment was a seed
experiment for the Giant RAdio Neutrino Detector (GRAND), see
e.g.~\citet{GRAND:WP}. The location considered is the crossing point of the
North-South and East-West arms of the 21\,CMA interferometer
(42.924211\,$^\mathrm{o}$\,N, 86.698273\,$^\mathrm{o}$\,E, 2\,534\,m). It is
indicated with a black dot in the middle of the right plot of
fig.~\ref{fig:map}.

The rock depth seen from the two selected view points was computed using a
\ics{turtle_stepper} object. The Chaîne des Puys area is described by a high
resolution local DEM with a squared grid of pad size $2.5\,\mathrm{m}$ and with
4401$\times$4401 nodes. The local map coordinates are given in Lambert~93
projection. The corresponding data set is represented on the left of
fig.~\ref{fig:map}. Outside of this area, a constant ground elevation or zero is
assumed. The rock depth through the Puy de Dôme is tracked from the view point
up to an altitude of 2\,000\,m, which is slightly above the highest summits of
the Massif Central (Puy de Sancy, 1\,886\,m). We scanned $601 \times 301$ lines
of sight with azimuth and elevation values spanning $60\,^\mathrm{o} \times
30\,^\mathrm{o}$, centered on the Puy de Dôme.

For the Tian Shan area, the Ulastai valley and its surrounding area is described
by 49 SRTMGL1\,\cite{SRTMGL1} tiles, ranging from 39\,$^\mathrm{o}$\,N to
46\,$^\mathrm{o}$\,N and 83\,$^\mathrm{o}$\,E to 90\,$^\mathrm{o}$\,E. Each tile
has 3601$\times$3601 nodes and the grid pad size is approximatively 30\,m.  The
corresponding data are represented on the right of fig.~\ref{fig:map}. Outside
of this area we assume a ground elevation of zero as well. The rock depth is
tracked from the view point up to an altitude of 7\,500\,m, which is slightly
above Jengish Chokusu peak (7\,439\,m), the highest summit of Tian Shan. We
scanned $1801 \times 241$ lines of sight with azimuth and elevation values
spanning $360\,^\mathrm{o} \times 12\,^\mathrm{o}$.

\section{Balancing speed and accuracy \label{sec:balancing}}

\begin{figure}[!t]
  \centering
  \subfloat{\includegraphics[width=0.5\textwidth]{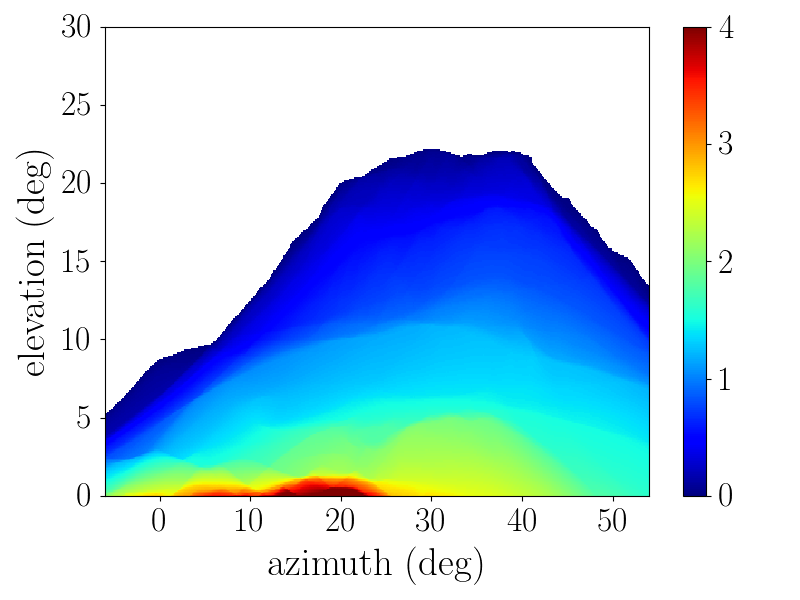}}
  \subfloat{\includegraphics[width=0.5\textwidth]{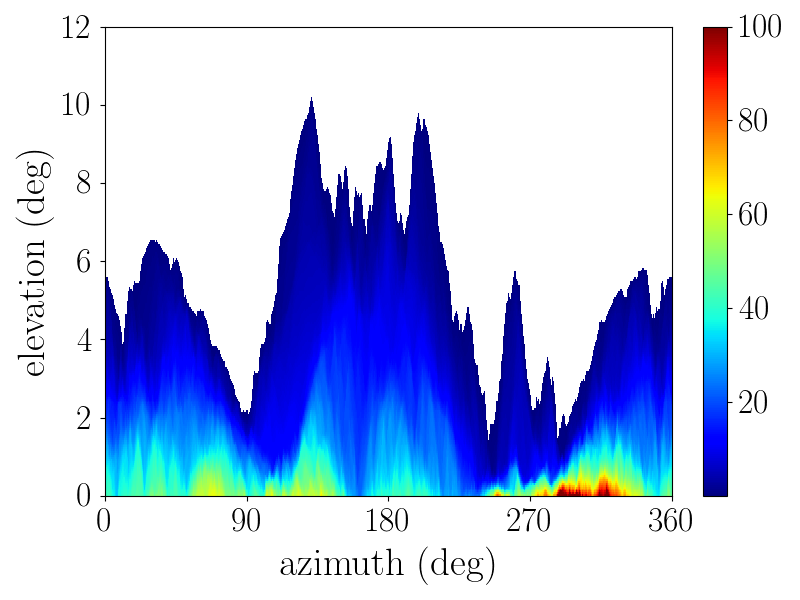}}
  \caption{Rock depth (km) as function of the direction of observation. The rock
        depth was computed with the reference method. Left: CDC view.  Right:
        Ulastai view.  An azimuth of $0\,^\mathrm{o}$ points to the geographic
        north and an azimuth of $90\,^\mathrm{o}$ points to the east.
        \label{fig:depth}}
\end{figure}

The \emph{optimistic} algorithm has three tuning parameters: the slope factor,
$\alpha$, the resolution factor, $s_\mathrm{min}$ and the binary search
accuracy, $\epsilon$. In addition, TURTLE introduces a linear approximation,
adding a fourth tuning parameter: the LLA range, $r_\mathrm{LLA}$. These four
parameters allow to balance accuracy versus speed when computing the rock
depth along a line of sight.

The accuracy of the binary search, $\epsilon$, can be set to rather small values
without any significant increase in computation time. This can be understood
since most steps do not cross a boundary, i.e.~do not trigger a binary search.
Furthermore, the convergence speed of the binary search goes as $1 /
\ln(\epsilon)$. In TURTLE, $\epsilon$ is set to a constant value of $10^{-8}\,$m
which is well below the typical accuracy of DEMs.

Estimating the accuracy of the computation is not as straightforward as it might
first seem. Systematically setting small step lengths results in the
accumulation of numerical rounding errors on large distances. For example,
incrementing the position by steps of 1\,cm results in a few mm deflection after
100\,km, using a 8~bytes \ics{double}. In the Ulastai view, this leads to
errors of several meters on the rock depth for some peculiar trajectories that
are grazing the ground. In the particular case of a straight line trajectory,
this can be solved by computing the step positions, $\vec{r}(s)$, from the line
parametric equation instead of incrementing them by 1\,cm, i.e.~as:
\begin{equation} \label{eq:line}
        \vec{r}(s) = \vec{r}_i + s \frac{\vec{r}_f - \vec{r}_i}{|\vec{r}_f - \vec{r}_i|}
\end{equation}
with $s$ the curvilinear abscissa (distance) of the step and $\vec{r}_i$
($\vec{r}_f$) the initial (final) position.  However, in the general case this
is not possible since the particle direction changes at each step.

For the present study, a \emph{reference} set of rock depths is computed with a
slope factor of $\alpha=1\,\%$ and a resolution factor of $s_\mathrm{min} =
1\,\mu\mathrm{m}$. No local approximation is used and the step position is
computed from equation~\ref{eq:line}. This \emph{reference} set is cross-checked
against rock depths obtained with a constant initial step length of $s_0=1\,$cm
and positions computed from equation~\ref{eq:line} as well. Note that the latter
computation requires 20 CPU-days in the case of the full Ulastai view, with
217,921 lines of sight.  Therefore, we do not use a step length lower than
1\,cm. The \emph{reference} case almost always produces identical results than
the $1\,$cm case, while being $\times$100 faster. In the few cases where
differences are observed, decreasing the initial step length to
$s_0=1\,\mathrm{mm}$ for these specific trajectories shows that the
\emph{reference} case is correct instead of the $s_0=1\,$cm computation. The
rock depths computed with the \emph{reference} case are shown in
fig.~\ref{fig:depth}. The typical value in the Chaîne des Puys area is a few km.
In contrast, in the Ulastai valley, the visible rock depth can be above 100\,km,
close to the horizontal.

\begin{figure}[!t]
  \centering
  \subfloat{\includegraphics[width=0.5\textwidth]{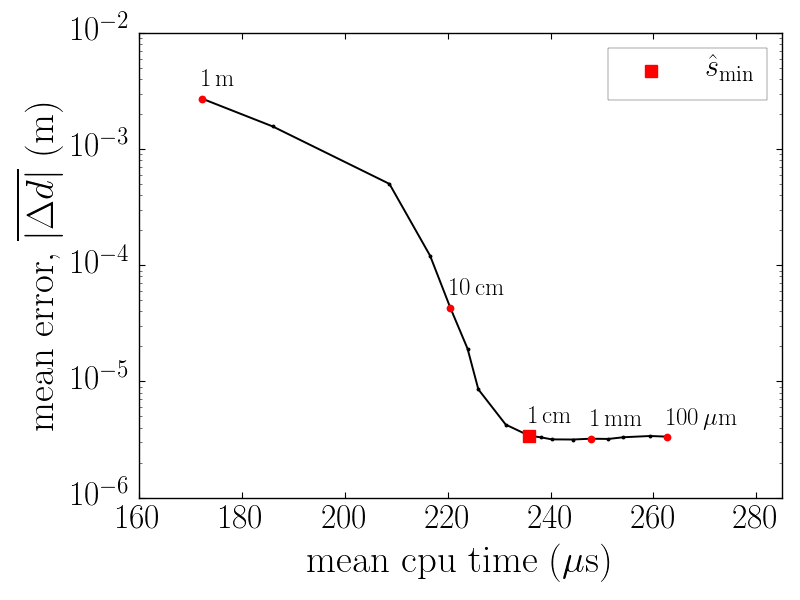}}
  \subfloat{\includegraphics[width=0.5\textwidth]{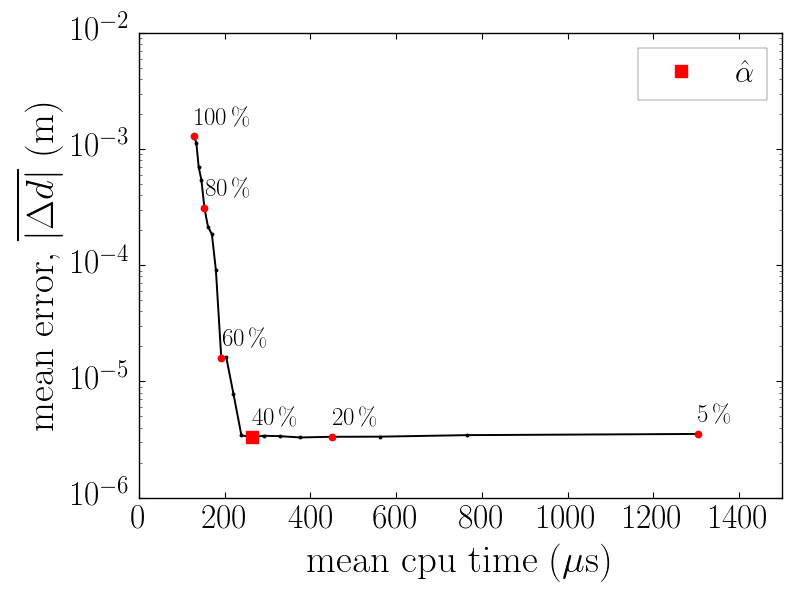}} \\
  \subfloat{\includegraphics[width=0.5\textwidth]{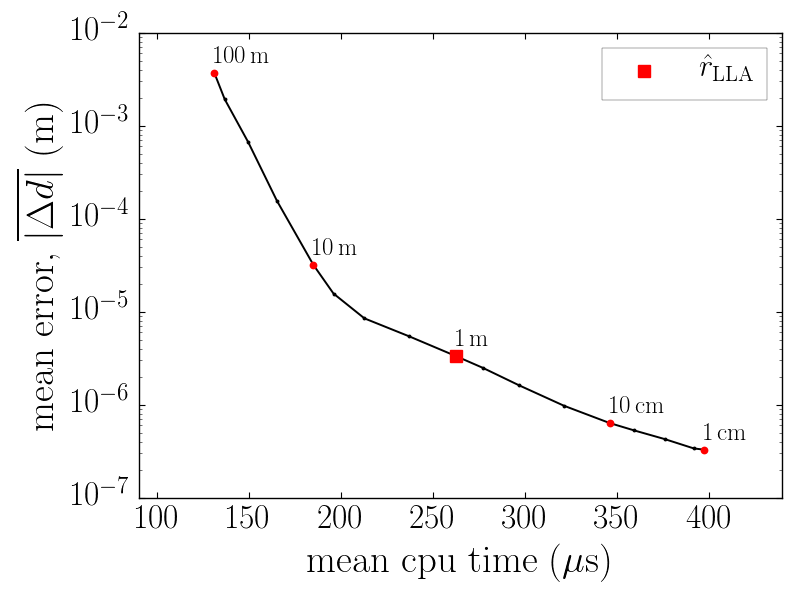}}
  \subfloat{\includegraphics[width=0.5\textwidth]{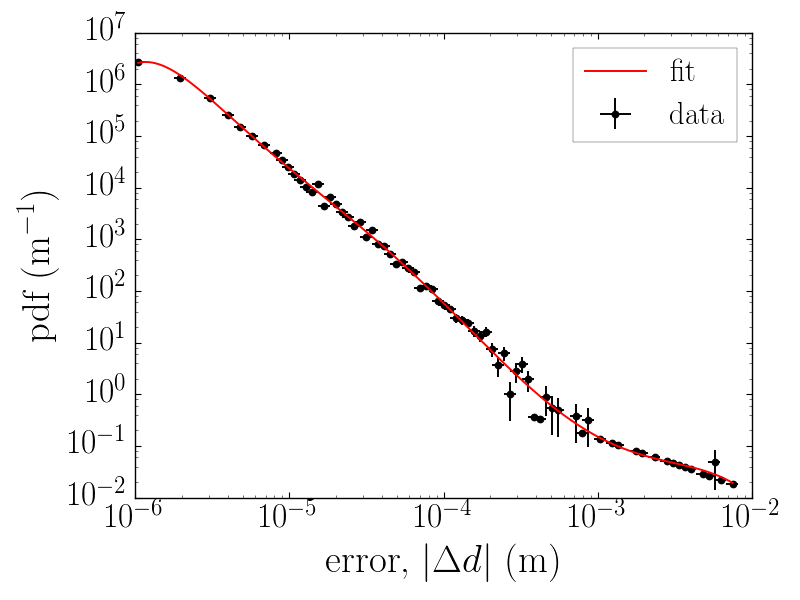}}
  \caption{Accuracy versus CPU time for the CDC view when moving the tuning
        parameters away from the selected values ($\alpha=40\,\%$,
        $s_\mathrm{min} = 1\,\mathrm{cm}$, $r_\mathrm{LLA} = 1\,\mathrm{m}$)
        indicated by a red square.  Top left: variation with the resolution
        factor. Top right: variation with the slope factor. Bottom left:
        variation with the LLA range. Bottom right: error distribution for the
        selected parameter values. The red line shows the result of a
        7$^\mathrm{th}$ order polynomial fit in log coordinates.
  \label{fig:tuning}}
\end{figure}

\begin{figure}[!t]
  \centering
  \subfloat{\includegraphics[width=0.5\textwidth]{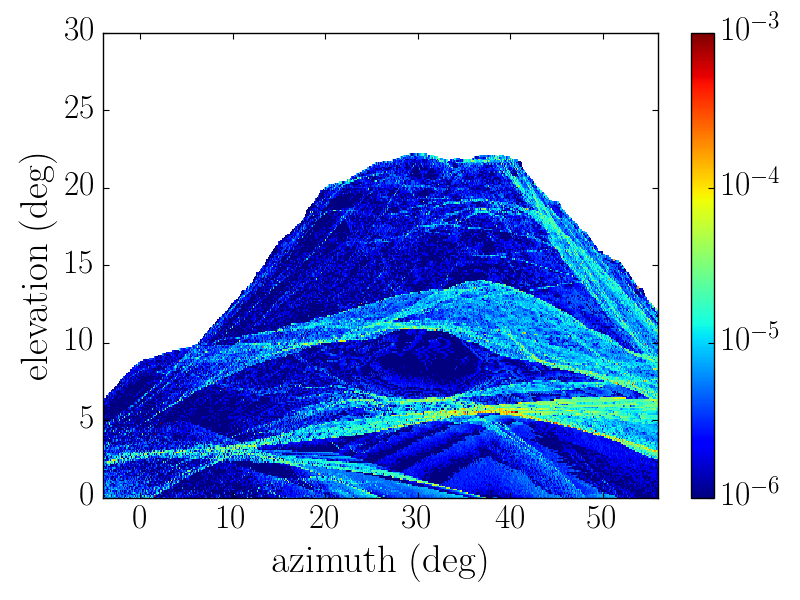}}
  \subfloat{\includegraphics[width=0.5\textwidth]{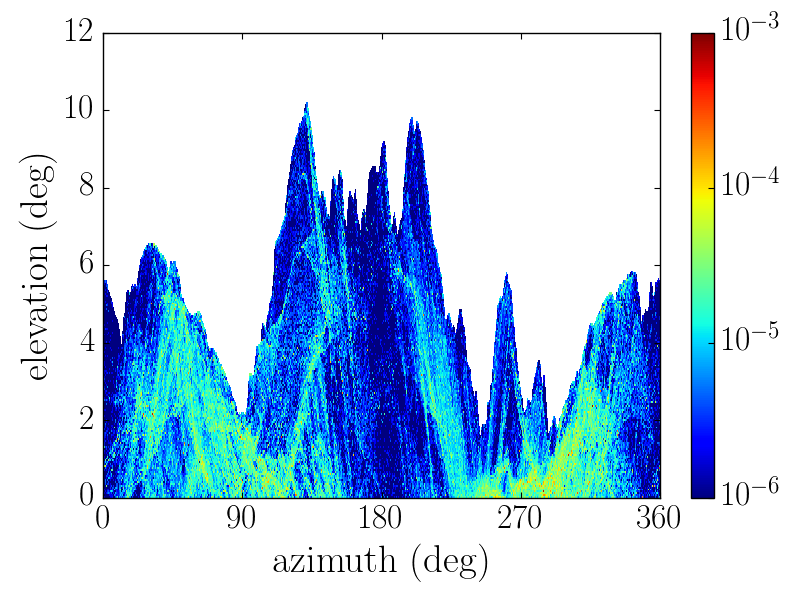}}
        \caption{Error on the rock depth (m) as function of the direction of
        observation. The rock depth was computed with the \emph{optimistic}
        algorithm using the selected parameter values: $\alpha=0.4$,
        $s_\mathrm{min}=1\,\mathrm{cm}$ and $r_\mathrm{LLA} = 1\,\mathrm{m}$.
        Left: CDC view. Right: Ulastai view. \label{fig:error}}
\end{figure}

\begin{figure}[!t]
  \centering
  \subfloat{\includegraphics[width=0.5\textwidth]{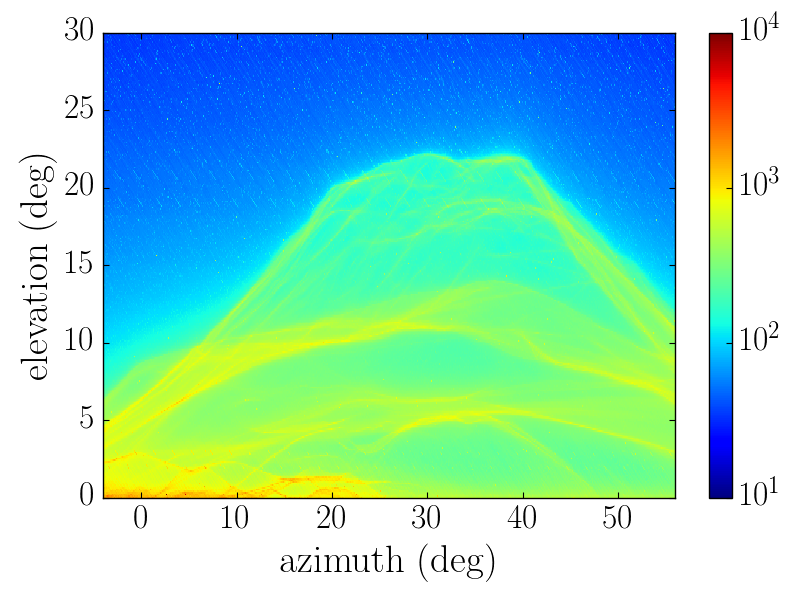}}
  \subfloat{\includegraphics[width=0.5\textwidth]{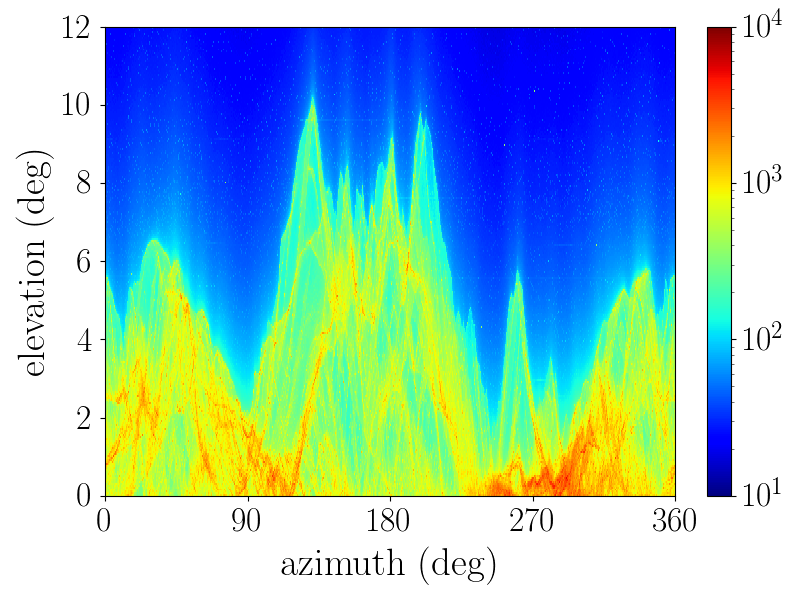}}
  \caption{CPU time ($\mu$s) for computing the rock depth as function of the
        direction of observation. The rock depth was computed with the
        \emph{optimistic} algorithm using the selected parameter values:
        $\alpha=0.4$, $s_\mathrm{min}=1\,\mathrm{cm}$ and $r_\mathrm{LLA} =
        1\,\mathrm{m}$. Left: CDC view. Right: Ulastai view.  \label{fig:CPU}}
\end{figure}

We scrutinize the parameters of the stepping algorithm. The rock depth for the
two views are computed for various parameter values using the
\ics{turtle_stepper_step} function. The track position is incremented step by
step as it would be in a MC. The absolute value of the difference, $|\Delta d|$,
w.r.t.~the \emph{reference} result is taken as an estimate of the overrunning
error on the rock depth, $d$. As a figure of merit we consider how the mean
error varies as function of the mean CPU time, averaging over all lines of sight
of a same view. Using this figure of merit, the following set of parameters has
been selected: $\alpha = 40\,\%$, $s_\mathrm{min} = 1\,\mathrm{cm}$ and
$r_\mathrm{LLA} = 1\,\mathrm{m}$, since it yields a good compromise between
speed and accuracy for the two views.  Figure~\ref{fig:tuning} shows how the
performances evolve when modifying the tuning parameters, one by one, around the
selected values. Figure~\ref{fig:tuning} corresponds to the CDC view, but
similar results are obtained for Ulastai one.  The selected parameters are
located close to a break point. Reducing them below the selected value results
in large CPU increase for little accuracy gain. For the selected parameter
values, it is also cross-checked that computing the steps positions from
equation~\ref{eq:line}, rather than incrementing them, does not significantly
reduce the error.

Figure~\ref{fig:error} shows the error on the rock depth, $|\Delta d|$, for the
selected parameter values. The mean of the error on the rock depth is low, 7 to
9\,$\mu$m depending on the view. The sign of the error is positive in $50\,\%$
of the cases, i.e.~there is an even probability to over or under estimate the
distance. This is a strong result, since it shows that the \emph{optimistic}
algorithm, though error prone, is not biased. The magnitude of the discrepancy
is only loosely correlated to the total rock depth: 5\,\% correlation factor for
the CDC view and 0.5\,\% for the Ulastai one. For the two views that we
consider, with the selected parameters, all errors are below $1\,$cm
\emph{except} for one line of sight in the Ulastai view. For an observation
angle of (55.4\,$^\mathrm{o}$\,N, 1.35\,$^\mathrm{o}$\,U) an error of 1.4\,m
occurs. Though, the relative error is of $49$\,ppm only. This problem is due to
a very sharp peak, located at ($43.093675\,^\mathrm{o}$\,N,
$87.038318\,^\mathrm{o}$\,E), for which the last meter before the summit is cut
through.  Reducing the slope parameter to $\alpha=0.2$ allows to resolve the top
of this peak for horizontal trajectories, but this comes at the cost of a 81\,\%
CPU increase.

A detailed look at individual tracks shows that the tail events, with errors of
several mm, are due to trajectories grazing the topography, i.e.~being almost
parallel to some parts of the ground. These effects are visible on the error
maps of Puy de Dôme (top plots of fig.~\ref{fig:error}). In particular, on the
left plot (CDC view) one clearly sees details of the Puy de Dôme surface due to
grazing rays. The top area of Puy de Dôme (above 10$\,^\mathrm{o}$) is very
steep from this side. The lines of sight make a large angle with the slope of
the volcano, and a good resolution is achieved. However, In the bottom area the
slope is milder. As a result the lines of sight graze the ground smoothing out
its roughness and leading to a poorer accuracy. It also explains the larger
correlation observed between $|\Delta d|$ and the total rock depth. The Ulastai
view is more complex since a single line of sight can include several features:
summits, valleys, etc. The absence of correlation between $|\Delta d|$ and the
total rock depth confirms that errors result from local peculiarities, e.g.~a
line of sight grazing a plateau, rather than from an accumulation of numerical
errors.

The consumed CPU time for each line of sight is shown in fig.~\ref{fig:CPU}.
Depending on the direction of observation, the CPU time varies by 2 to 3 orders
of magnitude from 10\,$\mu$s to 10\,ms. The CPU time increases abruptly when
crossing rocks. The lines of sight crossing rocks, half of the total number,
amount to 90\,\% of the CPU of a view. The CPU time is mainly driven by the rock
depth, with correlation factors of 73\,\% (Ulastai), and 87\,\% (CDC). The
secondary factor is the number of boundaries to map, i.e.~the complexity of the
topography. In addition, close to the horizontal, the CPU cost also increases
due to the very long paths needed in order to reach the exit altitude,
e.g.~$\sim$110\,km at $0\,^\mathrm{o}$ of elevation in order to reach 2\,000\,m
high, starting from Col de Ceyssat. The consumed CPU time is approximatively
proportional to the number of Monte~Carlo steps used in order to map a line of
sight. The average ratio is of 0.5\,$\mu$s per step for the CDC view.

\section{Comparison with ray tracing algorithms \label{sec:comparison}}

\begin{figure}[!h]
  \centering
  \subfloat{\includegraphics[width=0.5\textwidth]{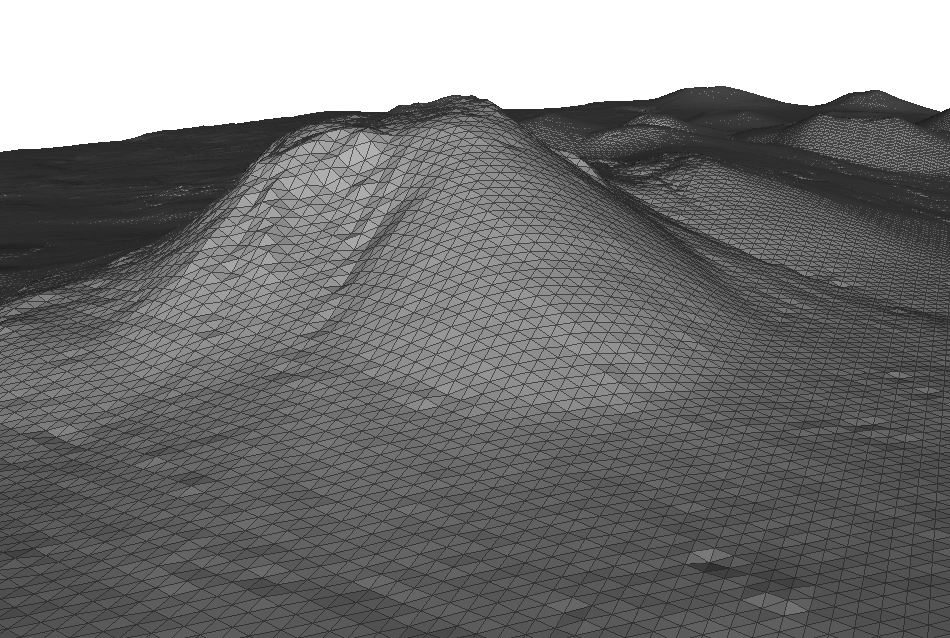}}
  \subfloat{\includegraphics[width=0.4\textwidth,trim=-1cm -1cm 0 0]{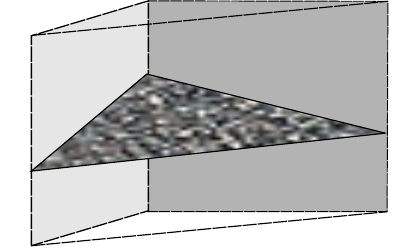}}
  \caption{Tessellation of the topography data for the BVH and polyhedral mesh
        algorithms. Left: tessellation of the ground surface using triangular
        facets joining the nodes. Right: schematic of a prism cell used for the
        tetrahedral mesh. The original triangular facet is the textured one.
        \label{fig:tessellation}}
\end{figure}

We compare the \emph{optimistic} algorithm discussed herein with the two
alternative ray tracing algorithms discussed in the introduction: a BVH tree and
a polyhedral mesh. In both cases the topography surface is modelled with
triangular facets, joining the data nodes, and using ECEF coordinates. The data
nodes of a same grid are regularly spaced in geodetic or projected coordinates.
Then, a square cell defined by four adjacent nodes is split in two triangular
facets, dividing the bottom right and the top left of the cell, as illustrated
on the left side of fig.\,\ref{fig:tessellation}. For the polyhedral mesh, the
geometry is divided in 3D cells using triangular prisms. For each triangle of
the tessellated topography surface we define a top (bottom) triangular prism,
having the triangular facet as bottom (top).  The sides of the prisms are
all parallel, defined by the local vertical at the middle of the topography. The
right side of fig.\,\ref{fig:tessellation} shows a schematic of the top and
bottom prisms of a triangular facet.

\begin{figure}[!h]
  \centering
  \subfloat{\includegraphics[width=0.5\textwidth]{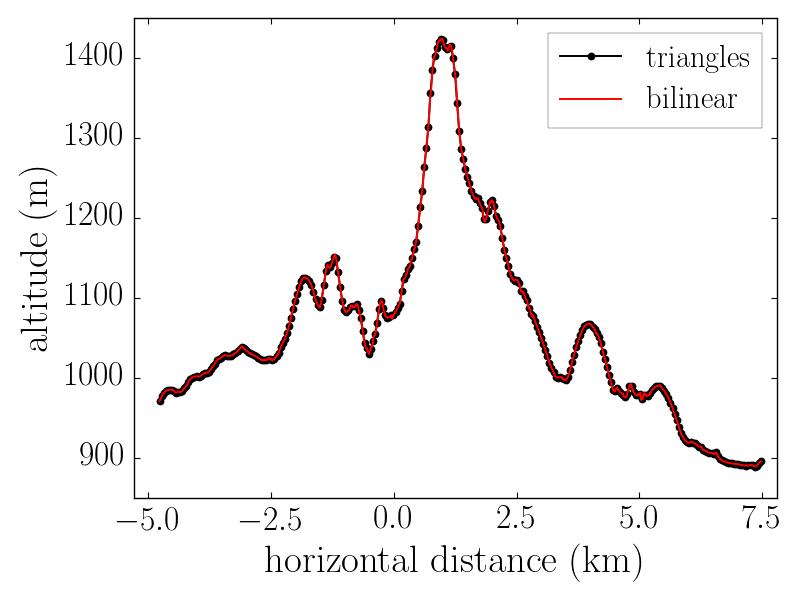}}
  \subfloat{\includegraphics[width=0.5\textwidth]{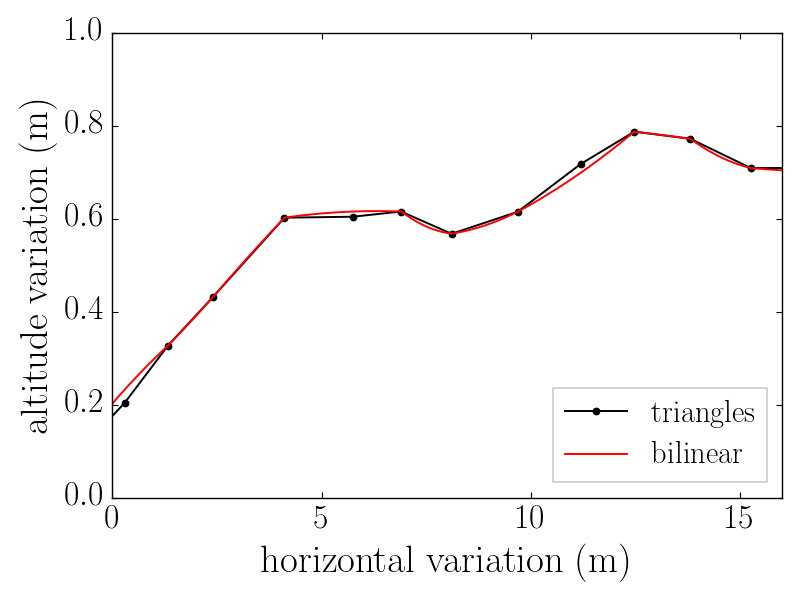}}
  \caption{Vertical section of the inner topography data at an azimuth of
        26$^\mathrm{o}$ from the CDC site. Red: bilinear interpolation (TURTLE).
        Black: triangular facets (BVH and polyhedral mesh). The dots indicate
        the intersection points with the borders of individuals facets.  Left:
        full section but showing only one out of 30 facets intersections.
        Right: zoom at a horizontal distance of 880\,m south of the CDC site.
        \label{fig:section}}
\end{figure}

Tessellating a topography surface with triangles is a classical representation
of a terrain. Let us recall that it differs from our approach which performs a
bilinear interpolation of the terrain between nodes. As an example
fig.\,\ref{fig:section} shows a vertical slice of the Chaîne des Puys topography
modelled with the two methods. On a large scale no difference is visible
(fig.\,\ref{fig:section}, left). However, when zooming down to the grid
resolution, discrepancies of several cm can be observed on the location of the
ground between two grid nodes (fig.\,\ref{fig:section}, right). For grazing
rays, this can result in differences on the rock depth of the order of the
spacing between grid nodes, i.e.~several meters in this case.

Both competitor algorithms are implemented using the CGAL
library\,\cite{CGAL:library} (v4.13), with a
\ics{CGAL::Simple_cartesian<double>} Kernel, as well as the ECEF conversion and
geographic projection routines from the TURTLE library. Note that for the BVH
method other libraries have been investigated as well, see e.g.~Embree\
\cite{Embree} in~\ref{sec:embree}.  The CGAL library was chosen because it
yields a good balance between speed and accuracy for ray tracing problems. For
the BVH method we use the generic AABB tree algorithm of~\citet{CGAL:AABB} with
\ics{CGAL::Triangle_3} primitives. The full geometry and its bounding boxes
needs to be instantiated resulting in a high memory overhead: 350 bytes per grid
node. Once initialised, the \ics{all_intersections} method of the
\ics{CGAL::AABB_tree} provide a fast computation of \emph{all} intersections of
the topography surface with a ray.  However, the intersections with the CGAL
AABB tree are not ordered w.r.t.~the ray origin.  Therefore getting the closest
intersection actually requires checking all intersections. Note that while the
\ics{CGAL::AABB_tree} has a \ics{first_intersection} method, it was found to be
significantly slower ($\sim$2$\times$) than looping over the result of
\ics{all_intersections} in order to find the closest one.

For the polyhedral mesh, a \ics{CGAL::Plane_3} object is used in order to
represent the prisms facets, positively oriented towards the outside of a cell.
The \ics{CGALL::Plane_3::has_on_positive_side} method allows to test if a
point lies on the inner or outer side of a facet. Since a triangular prism is a
convex volume, a point is located inside if and only if it is on the inner side
of all of its facets. Using triangular prisms instead of tetrahedra allows to
reduce the total number of facets to test when navigating through the geometry.
The total memory usage is kept low by creating and destroying the prisms on the
fly from the raw topography data.

The rock depth along a line of sight is computed following
pseudocode~\ref{alg:transport}. In the case of the BVH algorithm the
\ips{volume_at} function is implemented by counting the number of intersections
of a vertical ray with the topography surface. An even number of crossings
implies that the initial position is below the tessellated ground.  Then, the
\ips{distance_to} function proceeds by locating the closest intersection with a
ray along the line of sight. The intersection points define segments. The rock
depth is computed from the length of the segments lying below the surface. Note
that in this particular case where trajectories are straight, it is more
efficient to map all intersections with the BVH in a single step, since the CGAL
algorithm provides all of them by default. For the purpose of this comparison,
both methods are implemented for the BVH, i.e.~getting all intersections in a
single row, or stepping from one to another as in a MC.

\begin{figure}[!t]
  \centering
  \subfloat{\includegraphics[width=0.5\textwidth]{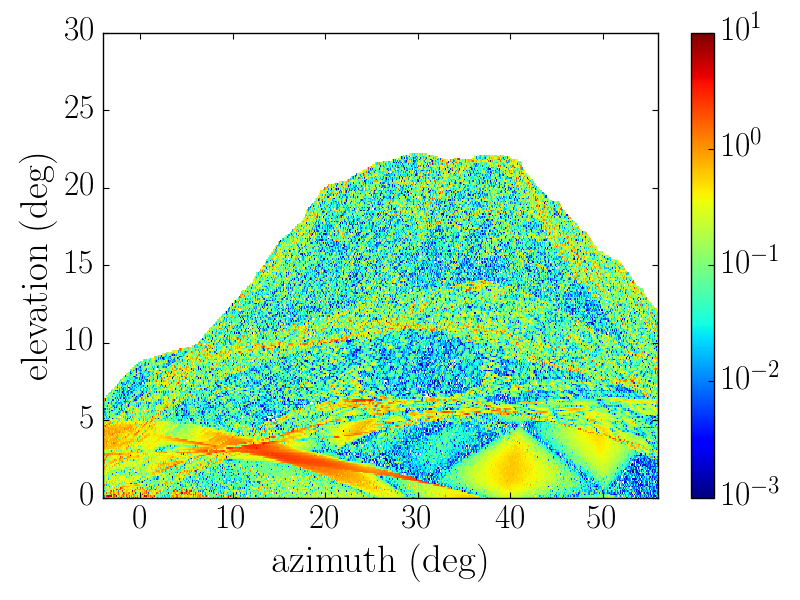}}
  \subfloat{\includegraphics[width=0.5\textwidth]{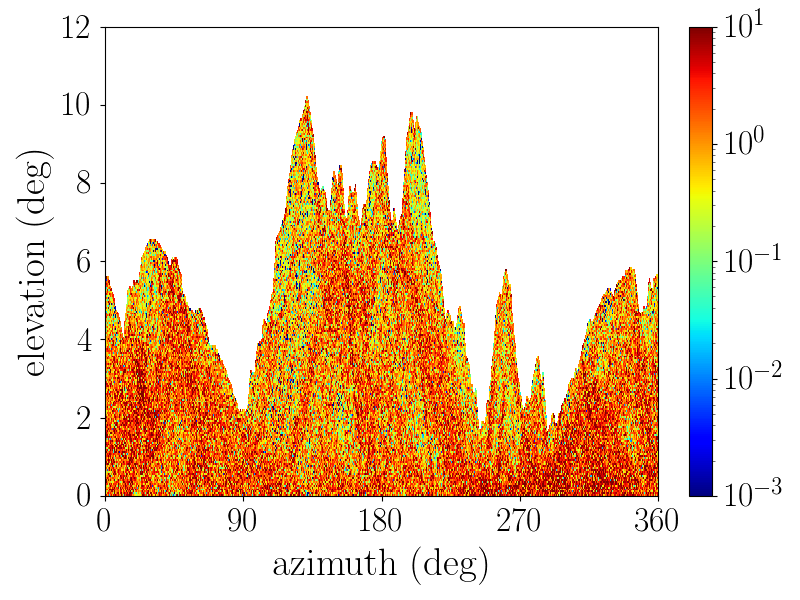}}
  \caption{Absolute differences on the rock depth (m), $|\Delta d|$, as function
        of the direction of observation, when interpolating the DEM with
        triangular facets or with a bilinear model.  Left: CDC view, right:
        Ulastai view. \label{fig:facets}}
\end{figure}

For the polyhedral mesh, the \ips{volume_at} function proceeds in two steps.
First we locate the closest grid node using geodetic or projected coordinates.
Then we test the prisms connected to this node. Once the initial volume is
located, we step through the prisms and locate the intersections with the
top or bottom facets. Note that since the prisms are connected by their facets
their is no need to re-locate the next prism when exiting the current one. The
exit facet determines the next sub-volume.

\begin{figure}[!t]
  \centering
  \subfloat{\includegraphics[width=0.5\textwidth]{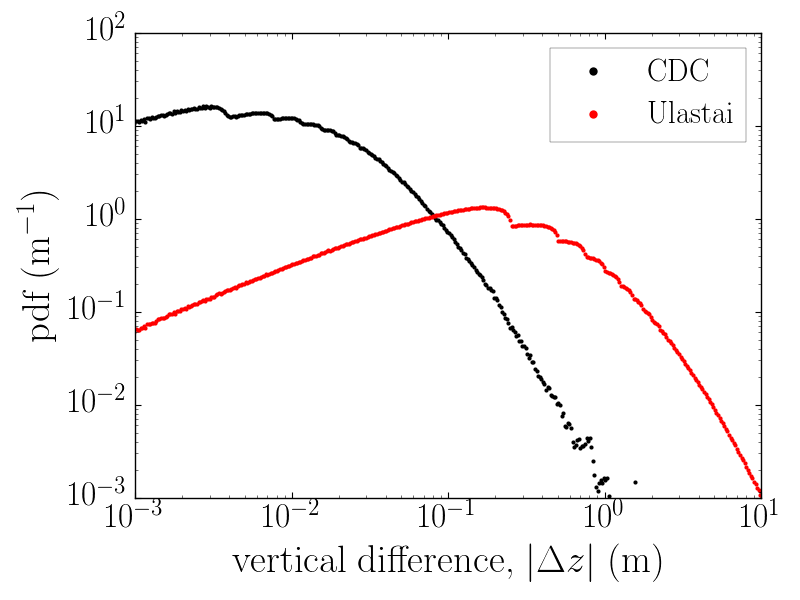}}
  \subfloat{\includegraphics[width=0.5\textwidth]{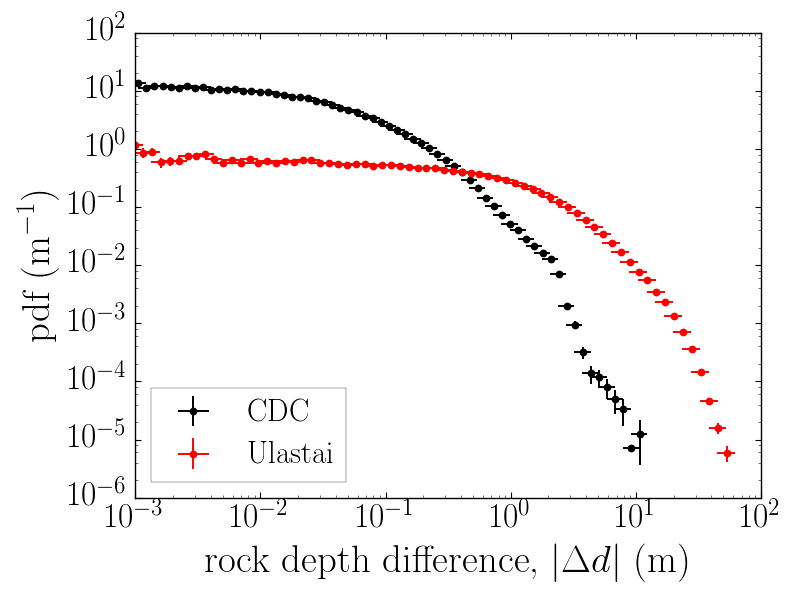}}
        \caption{Statistics of the modelling differences when using triangular
        facets or a bilinear interpolation.  Left: distribution of the absolute
        height difference, $|\Delta z|$. Right: distribution of the absolute
        difference on the rock depth, $|\Delta d|$.  \label{fig:facets_histo}}
\end{figure}

Fig.\,\ref{fig:facets} shows the absolute difference on the computed rock depth,
$|\Delta d|$, when using triangular facets or a bilinear interpolation of the
DEM data. A striking feature is that theses differences are much larger than the
overrunning errors of the tuned \emph{optimistic} algorithm
(fig.\,\ref{fig:error}). For the CDC view the average absolute difference on the
distance is 12\,cm. For the Ulastai view the average value is ten times higher:
118\,cm.  The fact that the differences are ten times larger for the Ulastai
view is consistent with the ten times larger grid spacing for the SRTMGL1 data
than in the inner grid used for the Chaîne des Puys.  It was also checked that
the signed differences, $\Delta d$, are evenly distributed in all three cases
with an average consistent with zero. The distribution of the absolute
difference on the rock depth, $|\Delta d|$, is shown on the right of
fig.\,\ref{fig:facets_histo}. For comparison, the distribution of the absolute
height difference $|\Delta z|$, is also shown on the left of the same
figure\,\ref{fig:facets_histo}. On average, those are approximatively five times
smaller than the differences on the distance: $\langle|\Delta z|\rangle=2\,$cm
for the CDC view and $19\,$cm for the Ulastai one.

\begin{figure}[!t]
  \centering
  \subfloat{\includegraphics[width=0.5\textwidth]{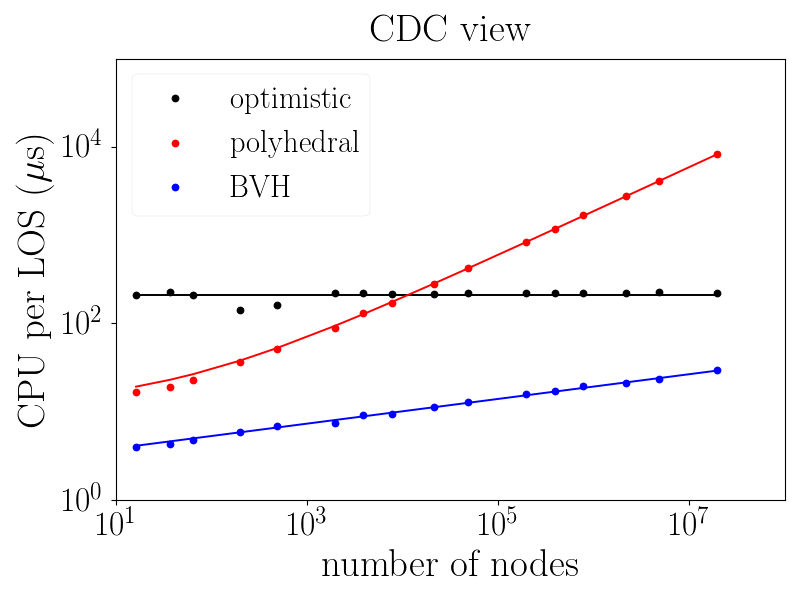}}
  \subfloat{\includegraphics[width=0.5\textwidth]{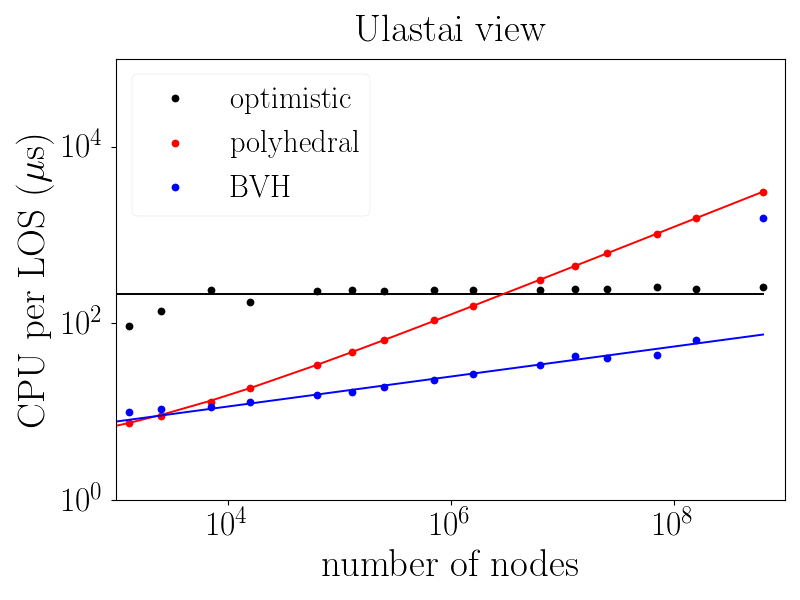}} \\
  \subfloat{\includegraphics[width=0.5\textwidth]{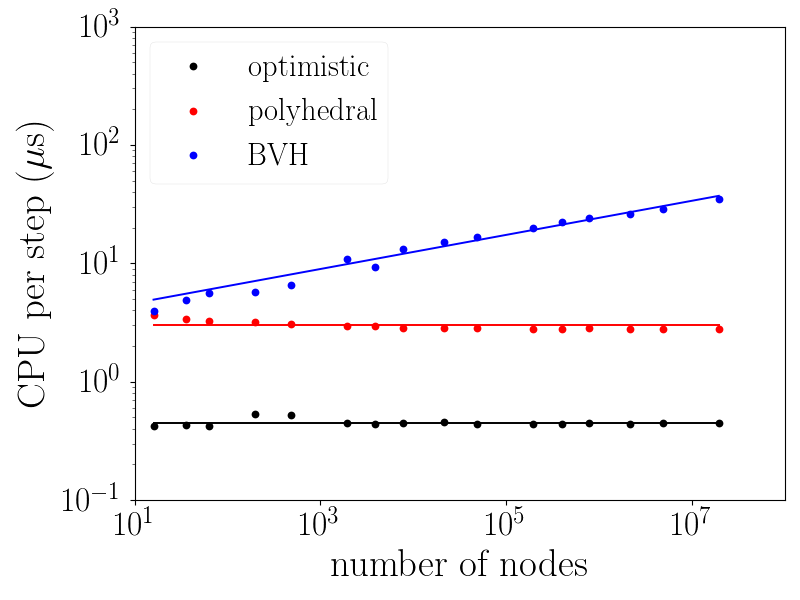}}
  \subfloat{\includegraphics[width=0.5\textwidth]{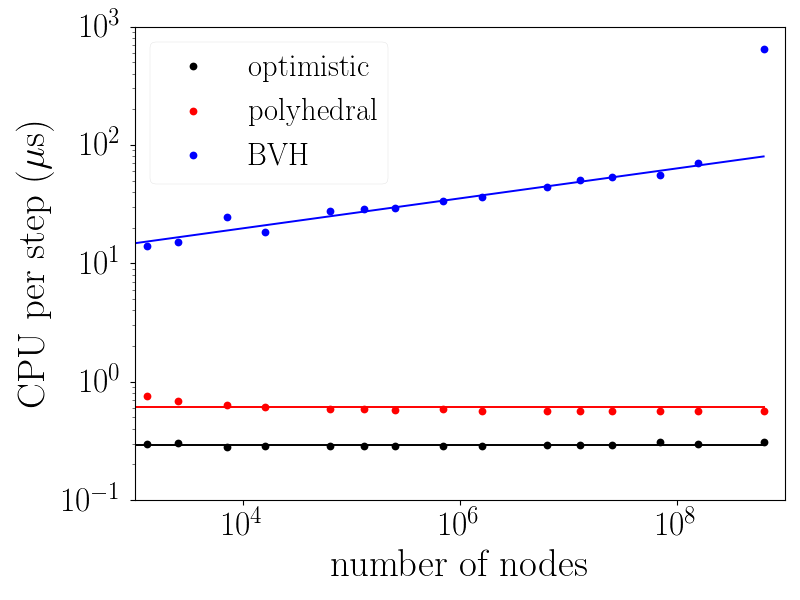}}
  \caption{CPU time ($\mu$s) as function of the number of grid nodes. Black:
        \emph{optimistic} algorithm. Red: polyhedral mesh. Blue: BVH tree. The
        solid lines correspond to the fit models described in the text. Upper:
        average CPU per line of sight for the CDC (left) and Ulastai (right)
        views. Bottom: CPU per Monte~Carlo iteration for the CDC (left) and
        Ulastai (right) views.  \label{fig:scaling}}
\end{figure}

The performances of the three algorithms are compared by varying the number of
nodes of the DEM data. A periodic down sampling is applied in order to vary the
number of nodes. We record the CPU time and the number of Monte~Carlo steps for
each line of sight. The upper part of fig.\,\ref{fig:scaling} shows the average
CPU time per line of sight as function of the number of nodes.  Numeric values
are also reported in table~\ref{tab:ray_tracing} for the case when no down
sampling is applied.  For the BVH algorithm intersections are computed in a
single row. Clearly, the BVH algorithm performs the best when it only comes to
map all intersections of the topography with a straight ray. The CPU time only
slightly increases with the number of nodes. The increase rate is found to
follow a power law: $a n^b$, with an exponent of $b = 1 / 8$.  However, for the
full Ulastai data 200\,GB of RAM are needed out of which the initial elevation
data represent only 1.2\,GB.  As a result the test server needs to use its swap
memory, resulting in a slowdown by a factor of ten when no down sampling is
applied. Furthermore, when the geometry has more than $n = 10^7$ nodes, the
initialisation of the BVH requires more time than scanning all the $\sim$200,000
lines of sight. The polyhedral mesh algorithm is not efficient for intersecting
large grids. Its CPU time is fitted with a square root law: $a + b \sqrt{n}$, in
agreement with expectations.  Interestingly, the performances of the
\emph{optimistic} algorithm do not depend on the number of nodes, but only on
the ray length and the topography features.  As a result it outperforms the
polyhedral mesh for grids larger than $1000 \times 1000$ nodes.  In principle it
should also be faster than the BVH algorithm for extreme grid sizes.  But, the
large amount of memory required by the BVH is likely to be an issue before this
happens. Let us recall that in contrast to a BVH, TURTLE's implementation of the
\emph{optimistic} algorithm requires no more memory than the initial elevation
data.

In the case where the step sizes are limited by the physics, the relevant
parameter is the average CPU time spent at each MC step in order to resolve the
geometry. An estimate of this quantity is provided by the total CPU time for all
lines of sight of a view divided by the total number of transport steps. In the
case of the BVH algorithm it is relevant to consider the CPU time with stepping
in this case. The CPU time per MC step is indicated on the lower part of
fig.\,\ref{fig:scaling} and in column four of table~\ref{tab:ray_tracing}. In
this configuration the BVH algorithm is inefficient. The \emph{optimistic}
algorithm performs best. It is a hundred times faster than the BVH algorithm,
for large grids. The polyhedral mesh is two (six) times slower than the
\emph{optimistic} algorithm, for the Ulastai (CDC) view. However, the present
comparison of the polyhedral versus \emph{optimistic} algorithm should be
considered with precaution. The CPU time per step can vary strongly in both
cases, depending on the step length dictated by the physics. In the polyhedral
case, the first step in a prism is much longer than the subsequent one, since
the local geometry needs to be built. A closer analysis, using
valgrind\,\cite{Valgrind:2007} \& callgrind\,\cite{Callgrind:2004}, shows that
building the prisms on the fly represents 75\% (Ulastai) to 93\% (CDC) of the
CPU instructions of a Monte~Carlo step, with our implementation. In the present
study each prism is traversed in a single step, which is the worst case.
Similarly, for the \emph{optimistic} algorithm, depending on the set LLA range,
successive small steps are accelerated by TURTLE.

\begin{table}[!h]
\begin{center}
\begin{tabular}{|c|c|c|cc|} \hline
    \multirow{2}{*}{View} & \multirow{2}{*}{Algorithm} & Steps   & \multicolumn{2}{|c|}{CPU ($\mu$s)} \\
                          &                            & per LOS & per LOS & per step \\
\hline \hline
        \multirow{3}{*}{CDC} & BVH        &    2.8 & 22 (51) & 18 \\
                             & Optimistic &    497 &     261 & 0.52 \\
                             & Polyhedral & 2\,975 &  8\,353 & 2.8 \\
\hline \hline
        \multirow{3}{*}{Ulastai} & BVH        &    4.3 & 46$^*$ (213$^*$) & 49$^*$ \\
                                 & Optimistic &    840 & 293    & 0.35 \\
                                 & Polyhedral & 5\,492 & 3\,233 & 0.59 \\
\hline
\end{tabular}
        \caption{Comparison of the performances of the BVH, \emph{optimistic}
        and polyhedral ray tracing algorithms for the 2 view points (CDC \&
        Ulastai) without down sampling. The CPU times are averages per line of
        sight or per transport step. The average number of transport steps per
        line of sight are reported as well. For the BVH algorithm the CPU time
        with and without stepping are quoted. $(^*)$\ For the Ulastai view with
        the BVH algorithm, due to memory issues, the quoted CPU times are
        actually extrapolations from a down sampled topography using the power
        law model described in the text. \label{tab:ray_tracing}}
\end{center}
\end{table}

Looking back at fig.\,\ref{fig:scaling}, or table~\ref{tab:ray_tracing}, one
might wonder why for the polyhedral mesh, the CPU time is five times larger in
the CDC view than in the Ulastai one.  A detailed analysis with callgrind shows
that this is due to the fact that the Chaîne des Puys inner map uses Lambert~93
projected coordinates, whose conversion has a high CPU cost: six times more
instructions than when using geodetic coordinates. As a result, building the
prism geometry is very inefficient. This could be solved by re-meshing the
elevation data in geodetic coordinates, or even better, directly in a plane
projection in ECEF coordinates.  The \emph{optimistic} algorithm uses a TURTLE
stepper with a 1\,m LLA range. It mitigates the CPU overhead introduced by the
map projection. As a result, the CPU time is only two times larger in the CDC
view than in the Ulastai one.

\section{Performances in Monte~Carlo simulations} \label{sec:hepmc}

When integrated in a MC, the ray tracing methods introduce a slowdown
w.r.t.~the pure physics processing. This slowdown arises from two factors.
First, computing the geometric step length adds an extra CPU cost at each
Monte~Carlo step. This factor is common to all geometry navigation methods. The
\emph{optimistic} and polyhedral mesh methods are designed in order to minimise
this cost. However, they add a second slowdown factor by increasing the total
number of Monte~Carlo steps required for the navigation through the geometry.
Depending on the CPU cost of each step, and the number of added steps, limiting
the geometric step length might be worth or not.

In order to check the performances of the TURTLE library, when integrated in a
MC, let us consider a classical problem for muography applications: the
computation of the atmospheric $\mu$ flux for each line of sight of the two
views previously considered. Such a problem is efficiently solved using a
reverse Monte~Carlo method, where the particles are tracked backward from the
detector to the source. Note that even though the Monte~Carlo flow is reversed,
the time flow is not. Reverse Monte~Carlo is a particular MC sampling method. It
does nor revert time neither solve an inverse problem. In the following we use
the words \emph{initial} and \emph{final} w.r.t.~time, i.e.~in their usual
meaning.  The PUMAS\,\cite{PUMAS:GitHub} library provides a reverse Monte~Carlo
engine for $\mu$ or $\tau$ implementing the \emph{backward} method. Details of
the backward method can be found in~\citet{PUMAS:2018}. For the present study,
version \texttt{0.14} of PUMAS is used, together with a \ics{turtle_stepper}.
PUMAS does not provide geometric primitives and ray tracing as for example
Geant4 does. Instead the PUMAS transport engine operates on a generic geometry
that must be supplied by the \emph{user} as a \ics{pumas_medium_cb} callback.
This callback must answer to \ics{volume_at} and \ics{distance_to} queries.
This allows to easily interface arbitrary ray tracers with PUMAS. An example of
ray tracer integration can be found in the source code of the test
suite~\cite{TURTLE_PERFS:GitHub} (function \ics{medium} in file
\ics{src/simulator.c}). Note that in order to avoid running superfluous ray
tracings, the result of the last \ics{distance_to} query is cached.

The TURTLE library and its \emph{optimistic} stepping scheme can also be
integrated in a generic Monte~Carlo like Geant4, without modifying the base
engine. This is further discussed in~\ref{sec:g4turtle}. However, for the
present study we decided to use a less complex Monte~Carlo engine, dedicated to
$\mu$ transport: PUMAS.

For each line of sight, the $\mu$ spectrum is estimated by backward sampling
100~$\mu$. The $\mu$ final kinetic energy is randomised over a $1 / E$
distribution between 1\,MeV and 1\,PeV. Note that \emph{a priori} any
distribution could be used for the final kinetic energy, as long as it covers
the full range of possible values. Using a $1 / E$ distribution is a good pick
for this problem, reducing the Monte~Carlo variance. It amounts to drawing the
logarithm of the final kinetic energy from a uniform distribution.  The particle
is then positioned at the view point, oriented backwards along the line of
sight.  From there on it is backward transported until it reaches the primary
source. The primary source is set at an altitude of 2\,000\,m (CDC) or 7\,500\,m
(Ulastai), as for the ray tracing problems. The maximum kinetic energy of the
primary $\mu$ source is set to 100\,PeV. Above this energy the atmospheric $\mu$
flux is too low, for muography applications. Whenever a backward propagated
$\mu$ exceeds this energy the tracking is stopped and its Monte~Carlo weight is
set to zero.

The PUMAS Monte~Carlo engine can be run with different levels of detail for the
physics configurable on the fly, e.g.~during the tracking of a particle. For the
present study we perform a \emph{detailed} simulation of physical processes.
The level of accuracy is comparable to Geant4 at low energy, below $\sim$1\,TeV,
and to dedicated high energy $\mu$ transport engines, e.g.~MUM\,\cite{MUM:2001},
above.

In a detailed simulation, due to scattering, particles observed along a given
line of sight from the view point actually follow different trajectories,
traversing different amounts of matter. This complicates performance studies.
Therefore, the PUMAS library is modified by overriding transverse deflections to
zero, such that all particles actually follow a straight trajectory along the
line of sight. Note that this modification is applied after computing any
deflection angle, in order to properly count the corresponding CPU cost.

The CPU cost of the physics simulation is estimated as following. Since all
particles follow the same straight trajectory, for a given line of sight, the
geometry can be pre-computed once for all, before actually running the MC
simulation.  This is done with TURTLE as well. The distances between the view
point and the intersections of the line of sight with the topography are stored
in a 1-dimensional table.  Then, during the MC simulation the distance of the
particle to the next boundary can be inferred from the table and its distance
from the view point. This can be done very efficiently such that the
corresponding CPU cost is negligible w.r.t.~the physics simulation. Running
PUMAS with this pre-computed geometry provides an estimate of the CPU cost of
the physics simulation.

The CPU cost for the physics simulation depends on the $\mu$ final energy. The
lower the $\mu$ energy the shorter the physical steps and the higher the
corresponding CPU cost. The CPU cost also increases in rocks since the physical
step length decreases by 3 orders of magnitude due to the differences in
target density. This is counterbalanced by the fact that the $\mu$ path length
in air, up to the primary altitude, is larger than in rocks.  Averaging over all
lines of sight and $\mu$ energies, the CPU cost is dominated by particles
traversing rocks and reaching the view point with low energy.

Table~\ref{tab:hepmc} shows a summary of the mean performances per line of sight
and per Monte~Carlo event for the two view points. Considering the physics only,
i.e.~when the geometry has been precomputed, on average $\sim\,312$ (380)
Monte~Carlo steps are needed per line of sight for the CDC (Ulastai) view. Most
of these MC steps concern the simulation of Coulomb multiple scattering. If this
process is disabled only a dozen MC steps are needed for the physics simulation.
In comparison, a pure ray tracing e.g.~with a BVH requires only 3 (4) steps for
the CDC (Ulastai) view (see table~\ref{tab:ray_tracing} in
section~\ref{sec:comparison}). This clearly shows that for the application
considered here most Monte~Carlo steps are limited by physics processes.

When the optimistic ray tracing is integrated in the MC simulation, the average
number of steps doubles for both views. One might notice that for the Ulastai
view the number of Monte~Carlo steps with ray tracing (797) is actually lower
than what was found previously (840) in section~\ref{sec:comparison}. This is
due to the fact that the backtracking is stopped when a $\mu$ energy exceeds
$100\,$PeV, which corresponds approximatively to the traversing of $\gtrsim
10\,$km of rocks. This seldom occurs in the CDC view where the largest rock
depth crossed is of $5.0\,$km. On the contrary, for the Ulastai view the largest
rock depth is of $117\,$km with an average value per line of sight of $6.7\,$km.

\begin{figure}[ht]
  \centering
  \subfloat{\includegraphics[width=0.5\textwidth]{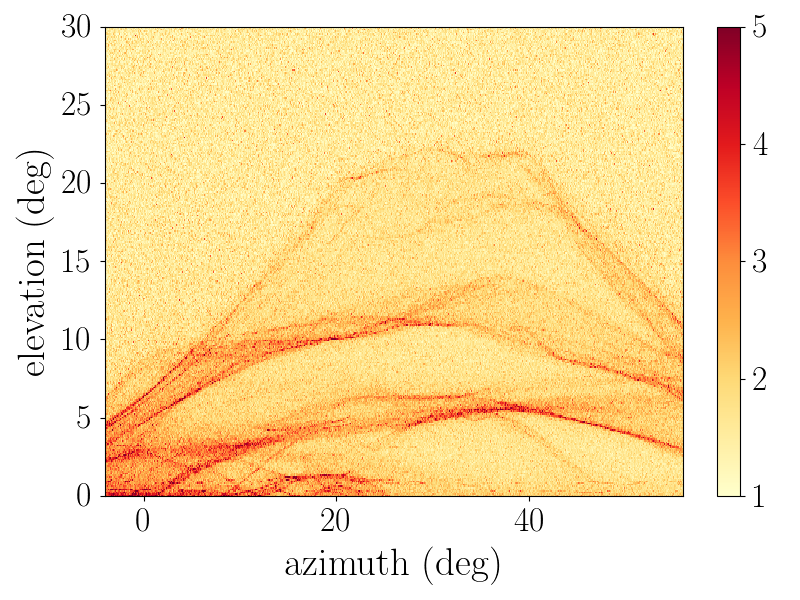}}
  \subfloat{\includegraphics[width=0.5\textwidth]{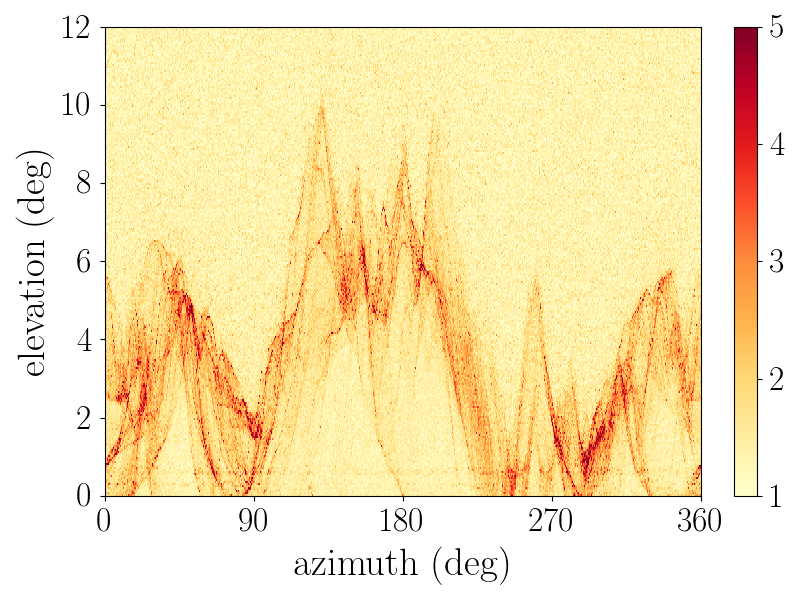}}
  \caption{Monte~Carlo slowdown factor as function of the direction of
        observation when computing a $\mu$ spectrum with PUMAS. The slowdown is
        defined as the ratio of the Monte~Carlo CPU with ray tracing to the one
        of the bare physics simulation. Left: CDC view, right: Ulastai view.
        \label{fig:detailed}}
\end{figure}

Figure~\ref{fig:detailed} shows the slowdown induced by the \emph{optimistic}
ray tracing, defined as the ratio of the CPU cost with ray tracing to the one of
the physics, i.e.~with a pre-computed geometry. Similar results are obtained for
both views. On average, the \emph{optimistic} ray tracing doubles the CPU cost
w.r.t.~the physics simulation. However, locally higher slowdowns (5-6) can be
observed for grazing rays. In such cases the number of Monte~Carlo steps
increases by one order of magnitude.  The average performances of the
\emph{optimistic} algorithm are very good given the large number of topography
nodes considered here. The cost of the geometry resolution is kept at the level
of the physics simulation. It implies that whatever the efficiency of a ray
tracer algorithm, at best it could speed up the present Monte~Carlo simulation
by a factor of two.
\begin{table}[ht]
\begin{center}
\begin{tabular}{|c|cc|cc|} \hline
        \multirow{2}{*}{View} & \multicolumn{2}{|c|}{Steps}    & \multicolumn{2}{|c|}{CPU (ms)} \\
                              & Pre-comp. & Optimistic & Pre-comp. & Optimistic \\
\hline \hline
        CDC     & 312 & 695 & 0.9 & 2.2 \\
        Ulastai & 380 & 797 & 1.0 & 2.2 \\
\hline
\end{tabular}
\end{center}
        \caption{Comparison of the Monte~Carlo performances using the
        \emph{optimistic} (TURTLE) ray tracing  or a pre-computed geometry for
        the navigation.  The results are given for the two view points (CDC \&
        Ulastai) without any down sampling.  The quoted CPU time and number of
        Monte~Carlo steps are averages per line of sight and per Monte~Carlo
        event.  \label{tab:hepmc}}
\end{table}

The Monte~Carlo performances of the \emph{optimistic} ray tracer have been
compared with the ones achieved when using the BVH and polyhedral mesh discussed
previously in section~\ref{sec:comparison}. Let us recall that the CGAL
library~\cite{CGAL:library, CGAL:AABB} is used for the BVH and polyhedral mesh
implementations. The exact same Monte~Carlo transport algorithm is used for all
ray tracers. Only the \ics{volume_at} and \ics{distance_to} callbacks differ.
The obtained numerical results are summarised in
table~\ref{tab:hepmc_comparison}. It can be seen that in this scenario where the
stepping is driven by the physics, the \emph{optimistic} algorithm (TURTLE)
performs the best.  It outperforms CGAL's BVH even for the CDC view where the
topography grid is of moderate size (~$\sim$$10^7$ nodes). For the Ulastai view
the quoted numbers for the BVH algorithm are underestimates.  Due to memory
issues, the DEM was down-sampled by a factor of two along both grid dimensions.
Otherwise, the required memory exceeds $200\,$GB resulting in a slowdown by a
factor of 10 due to swapping between the RAM and the hard drive. Note that the
CGAL BVH algorithm is particularly inefficient in this situation because it maps
\emph{all} intersections with the topography at each ray tracing query, instead
of just the closest one. More efficient BVH exist in this situation, as shown
in~\ref{sec:embree}. An optimised BVH ray tracer might improve the Monte~Carlo
CPU performances over TURTLE (the \emph{optimistic} algorithm), while providing
a similar accuracy. However, for large DEMs this would be at a high memory cost
($\sim$$100\,$GB) for a moderate CPU gain (a
factor of two at best).

The polyhedral mesh performs surprisingly well in the Monte~Carlo of the Ulastai
view. It is only two times slower than the \emph{optimistic} algorithm. This is
due to the fact that the backtracking is interrupted for long range particles,
i.e.~close to the horizontal, because the $\mu$ would be of too high energy as
discussed previously. This is reflected as well on the number of Monte~Carlo
steps which is significantly lower in the Monte~Carlo ($3\,331$) than for the
pure ray tracing ($5\,492$, see table~\ref{tab:ray_tracing}).  However, for the
CDC view the polyhedral mesh performs poorly because of the extra CPU cost due
to the coordinates conversion from Lambert projection to ECEF, as observed in
previous section~\ref{sec:comparison}. TURTLE mitigates this cost by using a
Local Linear Approximation (LLA) which makes the comparison with the polyhedral
mesh unfair in this case. Properly optimised the polyhedral mesh could be
competitive with the \emph{optimistic} algorithm when the stepping is driven by
the physics. The polyhedral mesh is however inefficient for pure ray tracing
problems over large DEMs, e.g.~transporting weakly interacting particles like
neutrinos. Therefore, it is less versatile than the \emph{optimistic}
algorithm.

\begin{table}[ht]
\begin{center}
\begin{tabular}{|c|c|c|cc|} \hline
        \multirow{2}{*}{View} & \multirow{2}{*}{Algorithm} & Steps   & \multicolumn{2}{|c|}{CPU} \\
                              &                            & per LOS & per LOS (ms) & per step ($\mu$s) \\
\hline \hline
        \multirow{3}{*}{CDC} & BVH        &    310 & 8.0 &  26 \\
                             & optimistic &    695 & 2.2 & 3.2 \\
                             & polyhedral & 3\,169 &  11 & 3.6 \\
\hline \hline
        \multirow{3}{*}{Ulastai} & BVH        &  > 378$^*$ & > 21$^*$ & > 59$^*$ \\
                                 & optimistic &    797 &  2.2 & 2.8 \\
                                 & polyhedral & 3\,331 &  5.0 & 1.5 \\
\hline
\end{tabular}
\end{center}
        \caption{Comparison of the Monte~Carlo performances using a BVH
        algorithm for the ray tracing, the \emph{optimistic} algorithm (TURTLE)
        or a polyhedral mesh. The average CPU times per line of sight and per
        Monte~Carlo step are reported as well as the average number of
        Monte~Carlo steps. $(^*)$\ For the BVH of the Ulastai topography the DEM
        was down-sampled by a factor of two along both grid dimensions due to
        memory issues. \label{tab:hepmc_comparison}}
\end{table}

\section{Conclusion}

The \emph{optimistic} algorithm is an efficient method for navigating through
topographies.  Contrary to traditional ray tracing methods, the
\emph{optimistic} algorithm proceeds by trials and errors. It takes the risk of
overrunning some details of the geometry. In the case of a topography described
by a DEM, this risk can be efficiently controlled by adapting the stepping
distance as a function of the height above ground. Using this strategy, errors
are kept below the native accuracy of DEM data. In addition, the
\emph{optimistic} algorithm naturally handles higher interpolation models
between the grid nodes than the traditional flat triangular facets. Differences
on the rock depth of the order of the distance between the grid nodes can be
observed, when using a triangular tessellation or a bilinear interpolation.

The TURTLE C library provides utilities for navigating through a topography
described by a DEM using the \emph{optimistic} algorithm. The elevation data of
a DEM are encapsulated in \ics{turtle_map} objects. Values between nodes are
rendered with a bilinear interpolation.  Collections of maps, as provided by
global DEMs for example, are managed with \ics{turtle_stack} objects. The
\ics{turtle_client} object provides thread safe access to stacks of maps. TURTLE
supports a few cartographic projections, including UTM. The TURTLE library also
provides transforms for representing the topography data in Cartesian ECEF
coordinates. The top level component of the library is the \ics{turtle_stepper}
object. It provides navigation functionalities.

The traversal time of the \emph{optimistic} algorithm does not depend on the
number of nodes of the DEM grid, contrary to other methods. It only depends on
the extent of the topography, and on its details. The more a ray is grazing the
ground, the longer its traversal time. In addition, the \emph{optimistic}
algorithm was implemented in TURTLE with zero extra memory cost, apart from the
initial DEM data.  Thus, this method performs particularly well for large grids,
with more than $10^9$ nodes. In such cases, while a BVH algorithm could be ten
times faster it would require hundreds of~GB of extra memory. Polyhedral meshes
are not competitive for large scales.

The \emph{optimistic} algorithm is particularly efficient when navigating
through a topography in detailed MC simulations. Applied to a muography
Monte~Carlo, the TURTLE library allows to render a large scale topography on the
fly while only slowing down the simulation by a factor of two, w.r.t.~the bare
physics.  This is due to the fact that the \emph{optimistic} algorithm provides
both fast estimates of the distance to the ground together with a fast traversal
time.  Fast estimates are required in order to efficiently track particles that
frequently change direction. BVH like optimisations are not well tailored for
such cases: large number of nodes ($>10^7$) but with a stepping limited by the
simulation of the physics. At best only moderate CPU gains could be achieved but
at the cost of a large memory usage.

The current implementation based on eq.~(\ref{eq:step_size}) could be further
refined.  A simple improvement would be to vary the slope parameter, $\alpha$,
depending on the map region. For example in plain regions, values of $\alpha$
larger than in mountainous areas could be used. This would however require a
preliminary analysis of the map data. The algorithm discussed in this paper is
already very efficient, despite being extremely simple. It has no memory
overhead and requires no initialisation, apart from loading the initial DEM
data.

\section*{Acknowledgements}
The authors thank two anonymous reviewers for their critical reading which
contributed to improve the present paper.  This research was financed by the
French Government Laboratory of Excellence initiative no. ANR-10-LABX-0006, the
Region Auvergne, the European Regional Development Fund and the France China
Particle Physics Laboratory. This is Laboratory of Excellence ClerVolc
contribution~n$^\mathrm{o}$~370.  The SRTMGL1 (v3) topographical data used in
this study were retrieved from the online USGS EarthExplorer and NASA Earthdata
Search tools, courtesy of the NASA EOSDIS Land Processes Distributed Active
Archive Center (LP DAAC), USGS/Earth Resources Observation and Science (EROS)
Center, Sioux Falls, South Dakota.

\appendix

\section{Comparison with Embree \label{sec:embree}}

The rock depth along a line of sight is computed following the
pseudocode~\ref{alg:transport} and using the Embree library\ \citep{Embree}
(version 3.5.2) from Intel. This library provides a highly optimised BVH
algorithm exploiting the Single Instruction Multiple Data (SIMD) capabilities
of CPUs. However, note that Embree uses a fixed single precision float (32
bits). On the contrary, CGAL allows arbitrary precision using templating. A
\ics{CGAL::Simple_cartesian<double>} Kernel was used for this work,
i.e.~double precision (64 bits).

As for the CGAL based algorithm, the \ips{volume_at} function is implemented by
counting the number of intersections of a vertical ray with the topography
surface. The \ips{distance_to} function is an encapsulation of Embree's
\ips{rtcIntersect1}, which directly provides the closest intersection of a ray
with the scene. The \ips{RTC_SCENE_FLAG_ROBUST} was set, forcing Embree to use
its most accurate ray tracing algorithm at the cost of downgraded CPU
performances. As for the CGAL implementation, we also implemented a
\emph{straight} version of the \ips{distance_to} function where all
intersections are mapped in a single call. With the CGAL library, using the
\emph{stepping} version for the ray tracing is two times slower than using the
straight one. However, with Embree the stepping version is only 14\,\% longer
than the straight one.

\begin{figure}[!t]
  \centering
  \subfloat{\includegraphics[width=0.5\textwidth]{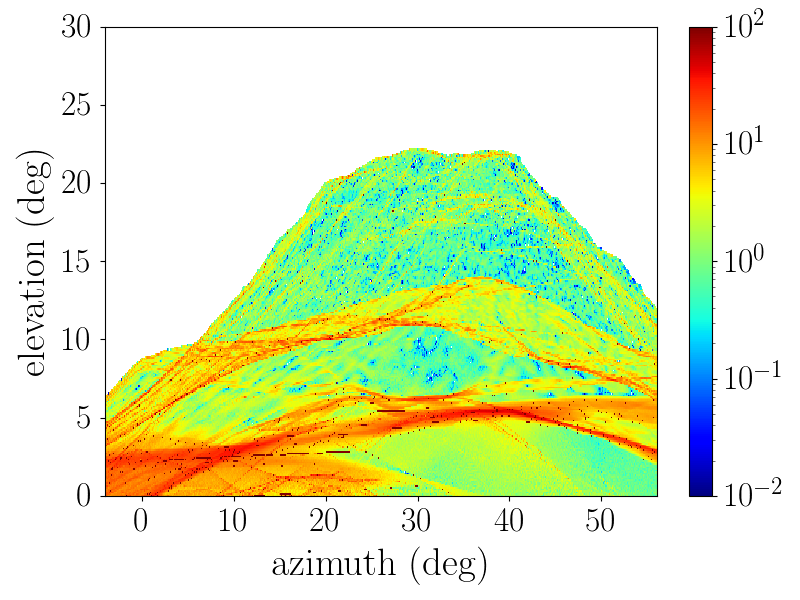}}
  \subfloat{\includegraphics[width=0.5\textwidth]{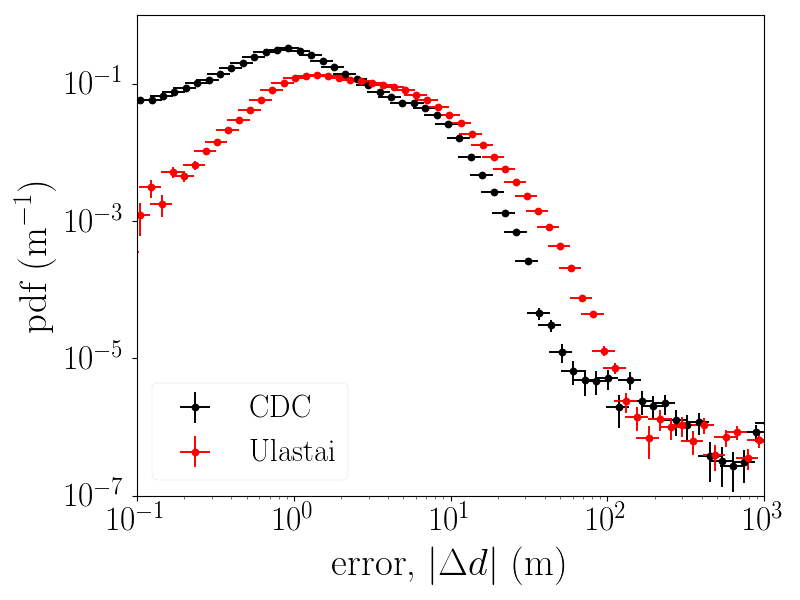}}
  \caption{Absolute error on the rock depth (m), $|\Delta d|$, using Embree.
        Left: error for the CDC view as function of the direction of
        observation.  Right: distribution of the error for the CDC (black) and
        Ulastai (red) views.  \label{fig:embree-error}}
\end{figure}

\begin{figure}[!t]
  \centering
  \subfloat{\includegraphics[width=0.5\textwidth]{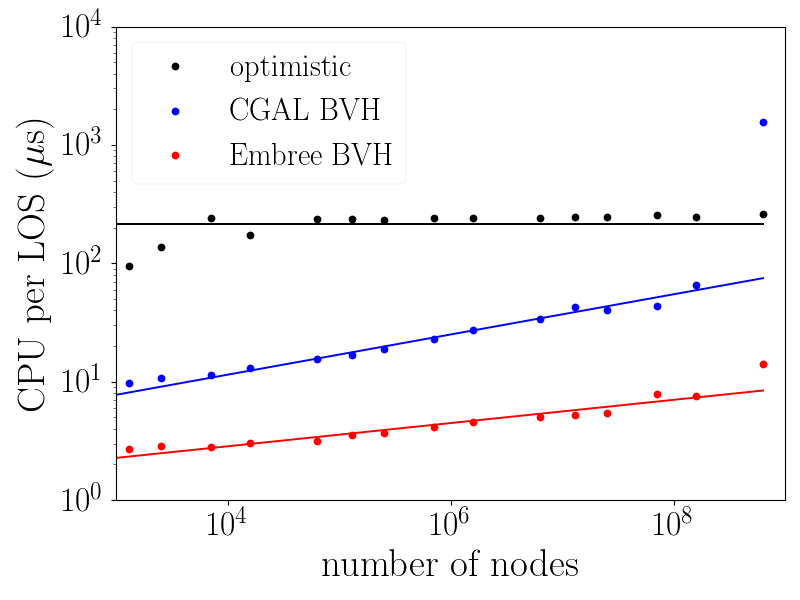}}
  \subfloat{\includegraphics[width=0.5\textwidth]{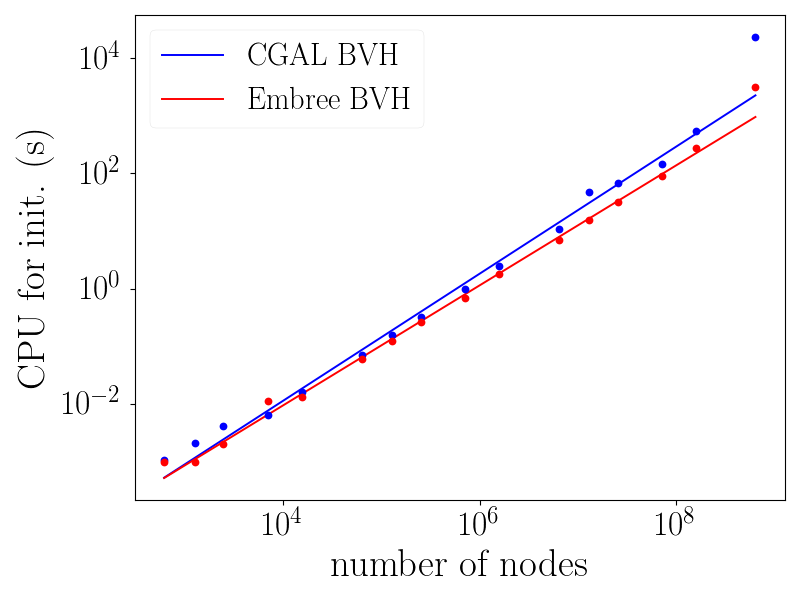}}
  \caption{CPU performances for the Ulastai view as function of the number of
        nodes.  Left: average CPU time ($\mu$s) of the ray tracing per line of
        sight. Right: CPU time (s) required for initialising the BVH. Black:
        optimistic algorithm. Red: Embree BVH.  Blue: CGAL BVH.
        \label{fig:embree-cpu}}
\end{figure}

Figures~\ref{fig:embree-error} and~\ref{fig:embree-cpu} show a selection of the
ray tracing performances obtained with Embree. A key issue is that Embree's
native numerical accuracy is insufficient for the physics application that we
consider. The ray tracing CPU performances of Embree are however impressive,
outperforming CGAL by one order of magnitude.  Note that the two libraries have
different scopes.  CGAL is a generic toolbox for applied Mathematics whereas
Embree is optimized for graphics rendering on CPUs.  Using Embree, the average
error on the rock depth is $4.3\ $m for the CDC view and $7.9\ $m for the
Ulastai one. Furthermore, errors as large as kilometers can occur along some
peculiar lines of sight. This clearly makes Embree not directly usable for most
physics applications. Note that Embree also allows custom geometries to be used.
This is done via user supplied callbacks implementing the bounding box
computation and ray intersection functions. It might be possible to improve the
numerical accuracy by using a custom double precision geometry with a slower but
more accurate math library.  Note however that Embree's internal representation
of bounding boxes would still be in single precision.  Implementing and
optimising this is beyond the scope of the present paper.

Despite its insufficient numerical accuracy, the Embree ray tracer was also
tested in the Monte~Carlo scenario of section~\ref{sec:hepmc}, i.e.~for a
muography application using PUMAS. The CPU performances are excellent with an
average slowdown of only 40\,\% w.r.t.~the physics simulation. This number could
be considered as a lower limit of the performances that could be achieved in
such a scenario. Note however that there is a trade to play between speed an
numerical accuracy such that an accurate enough implementation for physics
application is likely to be slower than what is achieved with Embree's native
ray tracer.
\begin{table}[ht]
\begin{center}
\begin{tabular}{|c|cc|cc|} \hline
        \multirow{2}{*}{View} & \multicolumn{2}{|c|}{Steps}    & \multicolumn{2}{|c|}{CPU (ms)} \\
                              & Pre-comp. & Embree & Pre-comp. & Embree \\
\hline \hline
        CDC     & 312 & 304 & 0.9 & 1.3 \\
        Ulastai & 380 & 377 & 1.0 & 1.4 \\
\hline
\end{tabular}
\end{center}
        \caption{Comparison of the Monte~Carlo performances using the
        \emph{Embree} ray tracer  or a pre-computed geometry for the 2 view
        points (CDC \& Ulastai) without down sampling.  The reported CPU time
        and number of Monte~Carlo steps are averages per line of sight and per
        Monte~Carlo event.  \label{tab:hepmc_embree}}
\end{table}

Finally, let us point out that the CPU time required to build the BVH scales
linearly with the number of grid nodes for both libraries. The BVH
initialisation requires $1.3\,\mu$s per node for Embree against $2.6\,\mu$s for
CGAL.  The memory usage of Embree is lower as well: 185 bytes per node. This is
approximatively half the requirements of CGAL which is consistent with the fact
that Embree uses single precision floats whereas double precision was used with
CGAL.

\section{Application to Geant4: G4Turtle \label{sec:g4turtle}}

The \ics{G4Turtle} class provides an example of interfacing of the TURTLE
library with Geant4. It encapsulates a \ics{turtle_stepper} object and its
related data : \ics{turtle_map}, \& \ics{turtle_stack}. The source code is
available from GitHub\,\cite{G4Turtle:Github} under the LGPL-3.0 license.  The
\ics{G4Turtle} interface is provided as a demonstration that the TURTLE library
can be integrated in a generic Monte~Carlo like Geant4, without modifying the
base engine. We do not claim that the present implementation is the most optimal
one. In particular, it does not support multi threading and it requires disabling
the \ics{SmartVoxels} navigation at the topography level.  It is functional
though, with a mild CPU slowdown per Monte~Carlo step. Below we provide an
overview of the \ics{G4Turtle} class followed by validation tests and
performance comparisons. The corresponding source code is available from the
test suite~\cite{TURTLE_PERFS:GitHub}.

\subsection{Description of the implementation}

In Geant4 a geometry is described by a set of volumes bounded by \ics{G4VSolid}
objects, e.g.~\ics{G4Box}, \ics{G4Orb}, $\ldots$ and filled with
\ics{G4Material}s. These volumes are ordered by inclusion relations. The top
volume is called \emph{World}. Each \ics{G4VSolid} implements in particular the
\ics{Inside}, \ics{DistanceToIn} and \ics{DistanceToOut} methods. These methods
allow to check if a particle is inside or outside of the volume, and to compute
the distance to the volume border, following a straight line along the particle
direction. Note that for these latter methods, Geant4 expects an exact distance
to be returned, given the particle direction. In order to implement the
\emph{optimistic} algorithm without modifying the \ics{G4Navigator}, a static
geometry must be mocked. In the following, we describe how this is done in
\ics{G4Turtle}.

A top level \ics{Envelope} volume is defined. This volume is bounded by a top
and bottom altitude, in ECEF coordinates. It contains two rock \ics{Chunks} and
two air \ics{Chunks} sub volumes. These sub-volumes are artefacts used to steer
the navigation. They have a variable position and extent. This required
disabling geometry optimisations within the \ics{Envelope} and its chunks, as
\ics{SetOptimisation(false)}. Otherwise the \ics{SmartVoxels} builder issues an
overlap exception. Geant4 navigation starts by checking if a particle is inside
the top volume, i.e.~the envelope. At this stage we compute the particle
altitude and the corresponding ground level using a \ics{turtle_stepper}. If the
particle is inside the Envelope, the Geant4 navigation will further check the
sub-volumes.  Given the particle altitude w.r.t.~the ground, the rock or air
chunks will return \ics{kInside} or \ics{kOutside}. Inside a \ics{Chunk}, Geant4
requests the distance to the border.  An \emph{optimistic} estimate of the
distance to the ground is returned, as given by
pseudocode~\ref{alg:optimistic:2}, i.e.~refined with a binary search. When the
navigation exits the chunk, it starts again checking if the particle is
still inside the envelope. However, it does not check the chunk that was just
exited.  Therefore, two chunks of each material are needed.

The previous method needs to be slightly refined in order to handle extra
volumes placed within the topography envelope, e.g.~a detector. First, when
Geant4 checks if the particle is inside the \ics{Envelope} we loop over all
extra sub-volumes, i.e.~non \ics{Chunks}. If the particle happens to be located
in any of these sub-volumes, the rock and air chunks return \ics{kOutside}, i.e.
extra volumes have precedence over the topography. In addition, when computing
the \emph{optimistic} stepping distance from within a \ics{Chunk}, we need to
check the \ics{DistanceToIn} to all extra sub-volumes. If any sub-volume is
closer than the distance provided by eq.~(\ref{eq:step_size}), the initial
stepping distance is modified accordingly.  If a binary search needs to be done,
extra sub-volumes are checked again, at each iteration.  Note that sub-volumes
are traversed with a linear search since \ics{SmartVoxels} navigation is
disabled. Therefore, this implementation is not optimal if a large number of sub
volumes is to be inserted inside the topography. \ics{SmartVoxels} are however
re-activated by default within sub-volumes, speeding up the navigation once
inside one of them. So, complex sub volumes, e.g.~a detailed detector, should be
manually enclosed in simple bounding \ics{G4Box} when placed within the
topography envelope.

\subsection{Validation and performances}

The \ics{G4Turtle} class is used with \ics{Geantinos} in order to compute the
rock depth for the two views described in section~\ref{sec:benches}. Version
10.5.1 of Geant4 is used. The results are compared to the one obtained with the
direct ray tracing of section~\ref{sec:comparison} using the \emph{optimistic}
algorithm. Note that both computations use a \ics{turtle_stepper} but wrapped
differently. An average agreement of a few $\mathrm{\mu m}$ is found on the rock
depth. While for almost all lines of sight the rock depth agrees better than
0.1~$\mathrm{\mu m}$, discrepancies up to 1\,cm can be observed for a few
grazing rays.

The number of steps required by the ray tracing is also in excellent agreement
for both cases, within $0.1\,\%$. Profiling with valgrind shows that
approximatively the same number of instructions are spent in the
\ics{turtle_stepper}.  However, the rock depth computation using Geant4
(\ics{G4Turtle} and \ics{Geantinos}) is 4 times slower than the dedicated ray
tracing.  This can be understood since Geant4 is a generic Monte~Carlo engine
not specifically optimised for a pure ray tracing problem. In this case most of
the CPU spent by Geant4 is for managing irrelevant functionalities for the ray
tracing, e.g.~\ics{G4Events}, \ics{G4Tracks}, etc.

\begin{figure}[!t]
  \centering
  \subfloat{\includegraphics[width=0.5\textwidth]{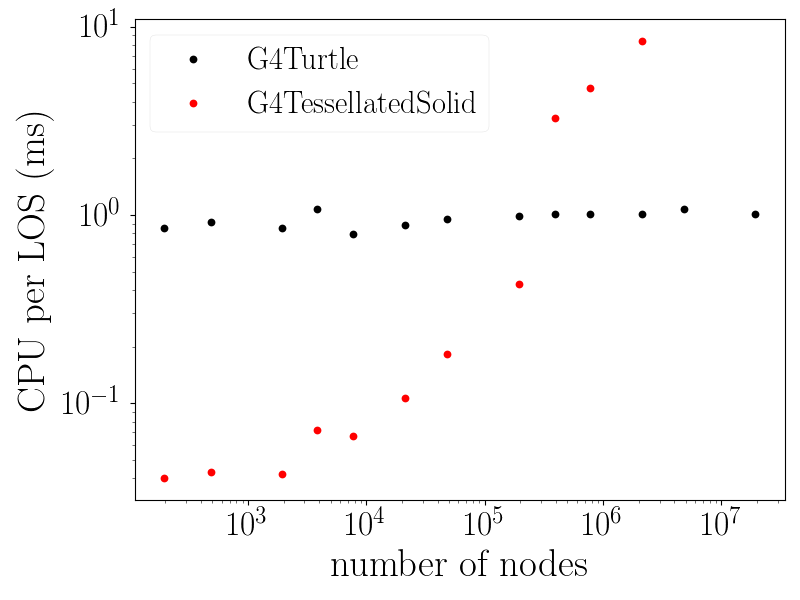}}
  \subfloat{\includegraphics[width=0.5\textwidth]{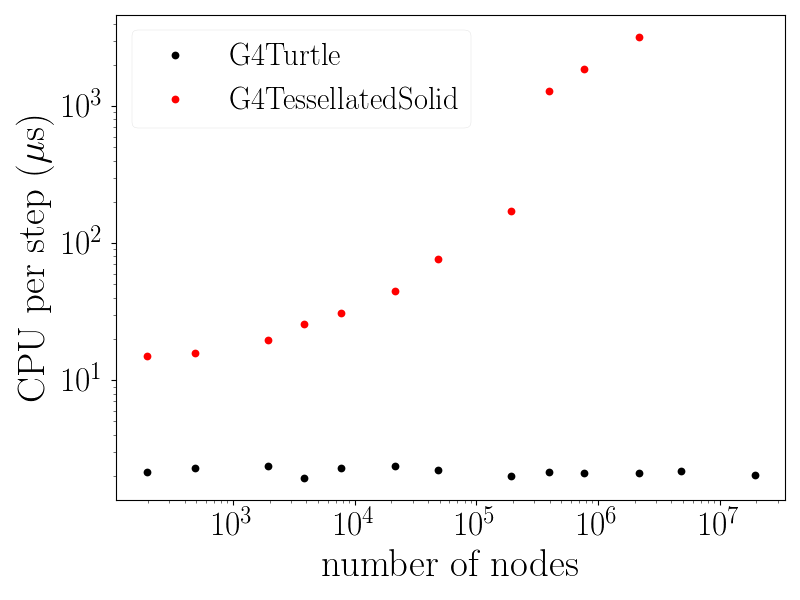}}
        \caption{CPU time as function of the number of grid nodes for a
        \ics{Geantino} scan of the rock depth seen from the CDC view point.  The
        (black) red dots stand for the \ics{G4Turtle} (\ics{G4TessellatedSolid})
        geometry.  Left: average CPU time per line of sight.  Right: average CPU
        time per Monte~Carlo step.  \label{fig:geant4-trace}}
\end{figure}

The performances of the \ics{G4Turtle} class have been compared to the one
obtained when modelling the terrain with Geant4's native
\ics{G4TessellatedSolid} geometry using \ics{G4TriagularFacets}. Only the CDC
view point was considered with various down-sampling values, as was done in
section~\ref{sec:comparison}. The initialisation time of the
\ics{G4TessellatedSolid} increases as the square of the number of nodes. Let us
recall that for CGAL and Embree's BVH the scaling is linear instead. This
prevents us for using the complete DEM data for the CDC view since the
initialisation would require approximatively two weeks.
Figure~\ref{fig:geant4-trace} shows the average CPU time per LOS and per step as
function of the number of nodes. For a moderate number of nodes ($\leq 10^4$)
the \ics{Geantino} scan is much more efficient with a \ics{G4TessellatedSolid}
than with a \ics{G4Turtle} geometry. However, above $10^4$ nodes the CPU time
per step required by the \ics{G4TessellatedSolid} starts to increase
approximatively linearly with the number of nodes. As a result, the
\ics{Geantino} scan becomes more efficient with a \ics{G4Turtle} geometry for
grids with $10^6$ nodes or more.

\begin{figure}[!t]
  \centering
  \subfloat{\includegraphics[width=0.5\textwidth]{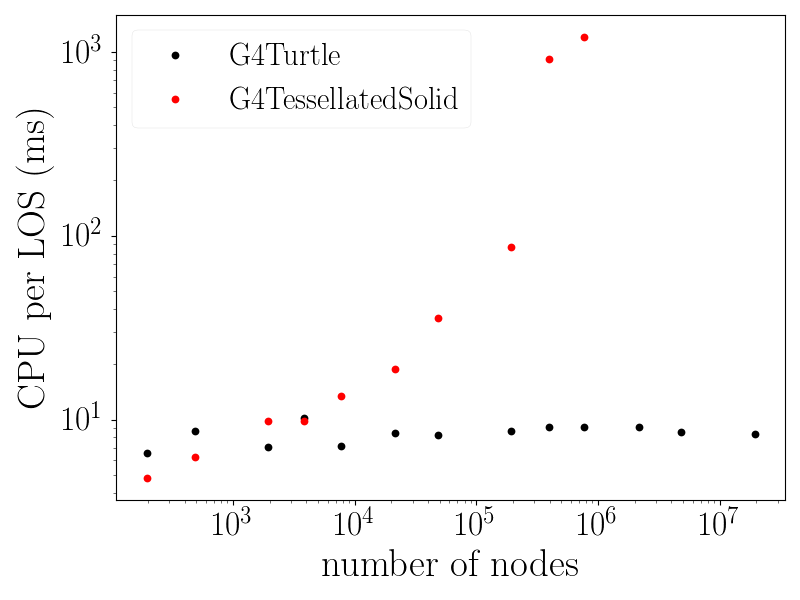}}
  \subfloat{\includegraphics[width=0.5\textwidth]{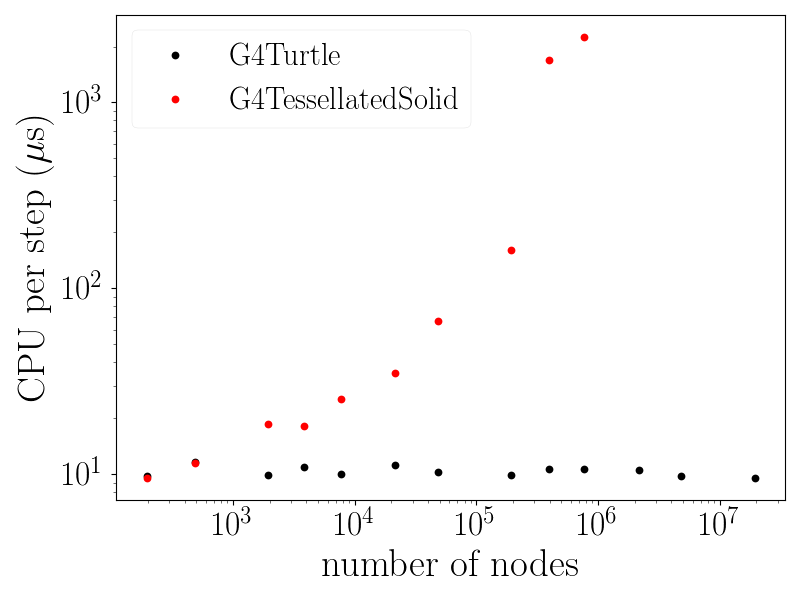}}
        \caption{CPU time as function of the number of grid nodes for the
        transport of 10\,TeV $\mu$ generated at the CDC view point.  The (black)
        red dots stand for the \ics{G4Turtle} (\ics{G4TessellatedSolid})
        geometry.  Left: average CPU time per line of sight.  Right: average CPU
        time per Monte~Carlo step.  \label{fig:geant4-sim}}
\end{figure}

Let us now use Geant4 for a pure $\mu$ transport problem similar to what has
been done in section~\ref{sec:hepmc} with PUMAS. High energy ($10\,$TeV) $\mu$
are generated at the viewpoint with an initial direction along a line of sight,
pointing upward. The $\mu$ is tracked until it escapes the simulation or decays.
The \ics{G4EmStandardphysics} and \ics{G4EmExtraphysics} modular physics list
are used and secondary tracks are killed, i.e.~only primary muons are tracked as
with PUMAS.  The total CPU cost depends strongly on the production cut value set
when building the \ics{PhysicsList}. For the present tests the cut value was
conveniently chosen to be of $1\,$m. For muography applications this would not
be accurate enough for lines of sights crossing a rock depth below
$\sim$$100\,$m. However, cut values below 1\,m significantly slow down the $\mu$
transport by increasing the number of Monte~Carlo steps.  The average CPU usage
per Monte~Carlo step does not significantly vary with the production
cut. The \ics{G4Navigator} amounts to 30\,\% of the CPU instructions
of a step out of which 3/4 are for the TURTLE library.

As for the \ics{Geantino} scan, the Monte~Carlo CPU performances of the
\ics{G4Turtle} geometry have been compared with the one of a
\ics{G4TessellatedSolid}. The results obtained by varying the number of DEM
nodes are shown on figure~\ref{fig:geant4-sim}. It can be seen that even for a
grid of moderate size ($\leq 10^4$) the \ics{G4Turtle} geometry is competitive.
For large grids it outperforms the current \ics{G4TessellatedSolid} because its
ray tracing becomes inefficient. From these numbers it can be seen that although
non optimal, the current implementation of \ics{G4Turtle} already delivers
excellent performances for stepping through topography data in Geant4.  However,
it does not allow to visualise the topography using one of the Geant4 3D
rendering modes.

Note that the VecGeom library~\cite{VecGeom:site} was investigated as well as a
replacement to the native \ics{G4TessellatedSolid} of Geant4. At the time of
this writing it suffers from the same flaws than the native Geant4
implementation. The initialisation time scales as the square of the number of
nodes and the ray tracing becomes inefficient for large grids. However, work is
ongoing in order to speed tessellated geometries in VecGeom by using the Embree
library. If this proves to be successful (accurate enough) then it would likely
outperform \ics{G4Turtle} for pure ray tracing problems, e.g.~\ics{Geantinos}
scan, even for large grids.  However, for most physical problems the Geant4 CPU
time would then be dominated by the simulation of the physics, especially for
small cut values. In such a case we would expect similar CPU performances for a
tessellation and \ics{G4Turtle}. Yet, \ics{G4Turtle} would require much less
memory for large grids.

\bibliographystyle{unsrtnat-url}
\bibliography{biblio}

\end{document}